\documentclass[preprint,superscriptaddress,preprintnumbers,amsmath,amssymb,prx,aps]{revtex4-2}

\usepackage{graphicx,ulem,mathptmx}

\usepackage{braket}

\begin{document}

\title{Continuous-variable quantum optics and resource theory for ultrafast semiconductor spectroscopy}

\author{Carolin L\"uders}
    \email{carolin.lueders@tu-dortmund.de}
	\affiliation{Experimentelle Physik 2, Technische Universit\"at Dortmund, D-44221 Dortmund, Germany}
 \author{Franziska Barkhausen}
	\affiliation{Department of Physics and Center for Optoelectronics and Photonics Paderborn (CeOPP), Universit\"at Paderborn, 33098 Paderborn, Germany}
\author{Matthias Pukrop}
	\affiliation{Department of Physics and Center for Optoelectronics and Photonics Paderborn (CeOPP), Universit\"at Paderborn, 33098 Paderborn, Germany}
\author{Elena Rozas}
	\affiliation{Experimentelle Physik 2, Technische Universit\"at Dortmund, D-44221 Dortmund, Germany}
\author{Jan Sperling}
    \email{jan.sperling@upb.de}
    \affiliation{Theoretical Quantum Science, Paderborn University, Warburger Stra\ss{}e 100, 33098 Paderborn, Germany}
   \affiliation{Institute for Photonic Quantum Systems (PhoQS), Paderborn University, Warburger Stra\ss{}e 100, 33098 Paderborn, Germany}
\author{Stefan Schumacher}
	\affiliation{Department of Physics and Center for Optoelectronics and Photonics Paderborn (CeOPP), Universit\"at Paderborn, 33098 Paderborn, Germany}
    \affiliation{Institute for Photonic Quantum Systems (PhoQS), Paderborn University, Warburger Stra\ss{}e 100, 33098 Paderborn, Germany}
	\affiliation{Wyant College of Optical Sciences, University of Arizona, Tucson, Arizona 85721, USA}
\author{Marc A\ss{}mann}
	\affiliation{Experimentelle Physik 2, Technische Universit\"at Dortmund, D-44221 Dortmund, Germany}



\begin{abstract} 
    In this review, we discuss the use of continuous variable spectroscopy techniques for investigating quantum coherence and light-matter interactions in semiconductor systems with ultrafast dynamics.
    We focus on multichannel homodyne detection as a powerful tool to measure the quantum coherence and the full density matrix of a polariton system.
    By monitoring the temporal decay of quantum coherence in the polariton condensate, we observe coherence times exceeding the nanosecond scale.
    Our findings, supported by proof-of-concept experiments and numerical simulations, demonstrate the enhanced resourcefulness of the produced system states for modern quantum protocols.
    The combination of tailored resource quantifiers and ultrafast spectroscopy techniques presented here paves the way for future applications of quantum information technologies.
\end{abstract}

\maketitle

\section{Introduction}

    Semiconductor systems offer a unique platform for implementing quantum technologies and quantum information applications.
    They can be engineered to show specific electronic and optical properties, and well-established and scalable standard manufacturing techniques enable their seamless integration into conventional technology and electronics.
    By combining semiconductor systems with nanostructures, it becomes possible to tailor their light-matter interaction, rendering them attractive for realizing integrated photonic circuits.
    The vast number of available materials has given rise to a wide range of semiconductor-based photonic integration platforms with different concepts for emitters, detectors, memories and other building blocks for optoelectronics and quantum technologies \cite{Pelucchi2022}.

    This rich variety of available semiconductor systems calls for a thorough way to characterize their light-matter interaction dynamics and their usefulness for applications in photonics and quantum information.
    The former task is the domain of semiconductor spectroscopy where the light-matter interaction is utilized to gain information about the material properties of the semiconductor itself.
    The latter question is investigated in quantum resource theories where the light-matter interaction is instead utilized to quantify the resourcefulness of the system for a specific task in quantum technologies.
    Still, in both cases, the light fields emitted from the semiconductor system carry the desired information.

    There are two fundamentally different ways to look at these light fields:
    The discrete-variable (DV) approach is more commonly applied in semiconductor physics and describes the light field in terms of the individual discrete photon number distributions of different modes and their photon number correlations.
    The typical detectors used in DV spectroscopy are photodiodes, superconducting nanowire detectors, as well as any other detector with photon-counting functionality.
    This is the natural approach to characterize, e.g., single-photon sources.

    In this review, we instead focus on the continuous-variable (CV) approach where each optical mode is considered as an individual electromagnetic harmonic oscillator, described by the quantum statistics of the continuous electromagnetic field's quadrature distributions of different modes and their correlations.
    Remarkably, in the quantum mechanical description of the light field, the two orthogonal field quadratures can be mapped to the position and momentum of a harmonic oscillator.
    This is the natural approach to describe quantum properties, such as phase and amplitude squeezing that are related to the fields themselves.
    The typical detection scheme for CV measurements is homodyne detection.

    Polaritons---hybrid light-matter quasiparticles arising from the strong coupling between excitons and microcavity photons---are ideal for a CV description.
    They may form a macroscopically populated condensate state that can be described by an order parameter in terms of an effective field.
    As leakage through the microcavity mirrors is the main decay mechanism of polaritons, the light field emitted from the cavity directly mirrors the properties of the polariton field.
    Recently, studies of quantum properties of polaritons have become a prominent research topic.
    For example, intensity squeezing of polaritons was demonstrated \cite{Boulier2014}. 
    Additionally, the observation of single polariton interactions \cite{Cuevas2018} and hydrodynamics \cite{Suarez2020} were reported.
    Integrated quantum polariton circuits were realized as well \cite{Nigro2022}.
    In this review, we are going to discuss how multichannel homodyne detection may be utilized as a powerful spectroscopy technique for ultrafast semiconductor physics and will demonstrate how it may be utilized to determine the quantum coherence and the full density matrix of a polariton system.

\section{Continuous variables and quasi-probabilities in phase space}

\begin{figure}
    \includegraphics[width=0.98\textwidth]{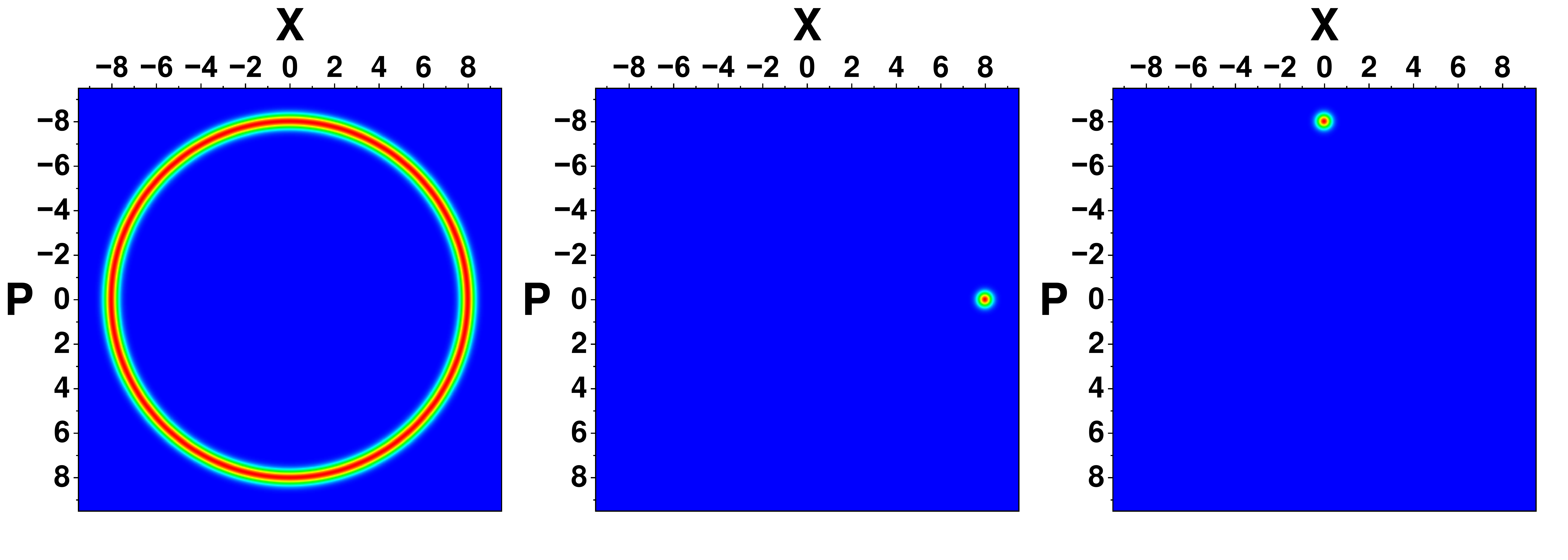}
    \caption{
        Phase space distribution for simultaneously measuring the normalized position $X$ and the normalized momentum $P$ of a classical harmonic oscillator.
        For measurements at random times, the probability distribution forms a ring with a width limited only by the resolution of the experimental setup (left panel).
        When measuring the system twice with a delay of a quarter period, both individual measurements still show rings, but correlations can be revealed.
        Filtering for all instances when the first measurement yields maximal positive displacement (center panel) always yields maximal negative momentum in the second measurement (right panel).
    }\label{fig:ClassicalPS}
\end{figure}

    For a classical harmonic oscillator, complete information about the system is always available in principle.
    Its position $x_\text{HO}$ and momentum $p_\text{HO}$ can be measured simultaneously up to arbitrary precision.
    One may then describe the harmonic oscillator in a phase-space picture by introducing a two-dimensional joint probability distribution for $x_\text{HO}$ and $p_\text{HO}$ which yields the relative probability to find a certain combination of $x_\text{HO}$ and $p_\text{HO}$ when measuring both observables simultaneously without any prior information.
    Considering an oscillator with unity mass, due to conservation of energy, the resulting distribution is an infinitely narrow circle in phase space; see Fig. \ref{fig:ClassicalPS}.

    In classical optics (CO), we can describe the light field in terms of a complex amplitude $\alpha$.
    We can consider the real and imaginary parts of the complex amplitude separately, $\alpha=\frac{x_\text{CO}+i p_\text{CO}}{\sqrt{2}}$.
    These form a set of orthogonal field quadratures and describe, e.g., the sine and the cosine component of a light field for a fixed frequency.
    In full analogy to the classical harmonic oscillator, we may introduce a joint probability distribution for joint measurements of $x_{CO}$ and $p_{CO}$.
    In the absence of any prior information, this again produces an infinitely narrow circle in phase space.

    In quantum mechanics, the analogy between quadratures and the harmonic oscillator persists.
    In a quantum description, the quadratures can be expressed in terms of the bosonic ladder operators $\hat{a}^\dagger$ and $\hat{a}$ as follows:
    \begin{equation}
        \hat{x}=\frac{1}{\sqrt{2}}(\hat{a}^\dagger+\hat{a})
        \text{ and }
        \hat{p}=\frac{i}{\sqrt{2}}(\hat{a}^\dagger-\hat{a}).
    \end{equation}
    The reference axes are, in principle, arbitrary.
    In practice, however, one defines the quadratures relative to a reference phase $\phi$:
    \begin{equation}
        \label{eq:phasequad}
        \hat{x}_\phi=\hat{x}\cos(\phi)+\hat{p}\sin(\phi)=\frac{1}{\sqrt{2}}(\hat{a}^\dagger e^{i\phi}+\hat{a}e^{-i\phi}),
    \end{equation}
    so that $\hat{x}_0=\hat{x}$ and $\hat{x}_{\frac{\pi}{2}}=\hat{p}$.
    These two orthogonal quadratures are complementary observables obeying the commutation relation $[\hat{x},\hat{p}]=i$.
    In quantum mechanics, uncertainty relations prevent us from obtaining precise simultaneous knowledge of complementary variables.
    It is therefore not possible to define a joint probability distribution for complementary quantum mechanical observables.
    Still, it is possible to define phase-space distributions that can be used to calculate expectation values of the relevant quantum mechanical observables.
    This transition to the quantum domain comes at the price of introducing phase-space quasi-probability distributions that may become negative or highly singular and, thus, cannot directly be interpreted as proper probability distributions.
    The most common examples are the Wigner function, the Glauber-Sudarshan $P$-distribution and the Husimi $Q$-function, which we discuss in more detail later.

\subsection{Dynamics and conditional measurements}

    So far, we considered phase space distributions for individual measurements in the absence of prior information about the system.
    Considering again the simplest case of a classical harmonic oscillator, this means that the amplitude and the phase of the harmonic oscillator are unknown a priori.
    Such individual measurements do not provide information about the dynamics of the system.
    However, sought-after dynamical information may be gained by consecutive measurements on the same system that are separated by short delays.
    Consider again Fig. \ref{fig:ClassicalPS}, which shows the probability distribution for position and momentum of a classical harmonic oscillator.
    We assume that we perform two simultaneous measurements of position and momentum instead, which are performed with a delay that corresponds exactly to a quarter period of the oscillator.
    When considered on their own, both measurements will yield infinitely narrow circles in phase space, but the correlations between the measurements provide information about the desired dynamics:
    Whenever the first measurement finds the oscillator at maximum displacement and zero momentum, the second measurement will show maximum momentum and zero displacement.
    By varying the delay, one can easily identify the resonance frequency of the harmonic oscillator.

    Real physical systems are usually more sophisticated than a harmonic oscillator and may show nonlinearities and phase and amplitude fluctuations.
    A standard way to study the dynamics of an arbitrary system is firstly to prepare it in a certain combination of position and momentum and then to measure the system dynamics following from this initial state; repeat this procedure for each possible combination of initial positions and momenta. 
    Instead, information about the system dynamics may also be gained without any preparation procedure by the consecutive measurement scheme outlined above.
    When we perform a series of joint measurements of position and momentum at varying delays, we may filter the joint results for all instances of a certain combination of position and momentum occurring in the first measurement.
    This conditional time-resolved measurement yields similar information as the time-resolved measurement of a system prepared with the same momentum and position.
    This postselection procedure has a clear drawback.
    It is less controllable, and one has to wait until some certain combination of position and momentum of interest actually occurs.
    However, it also has several advantages.
    Most importantly, if all measurement results are recorded individually, the combination of position and momentum that is of interest may be chosen after the measurement has already been performed.
    Also, postselection may be performed on both measurement results so that it is possible to investigate the dynamics at negative delays.
    Where preparation only provides experimental access to the dynamics after the preparation, postselection can also yield the dynamics before the system was in the state of interest.
    Also, not every system is fully predictable, and, for example, for open systems, rare events due to coupling to the surroundings of the system can occur.
    Consider, e.g., a noise-induced switching process in a bistable system.
    Such rare, random events usually cannot be prepared but will show up in postselection.
    It would be highly interesting to perform similar postselective measurements in phase space also in the quantum domain and in ultrafast semiconductor spectroscopy.
    In the following, we show how multichannel homodyne spectroscopy can achieve exactly this multilayered task.

\subsection{Homodyne detection}

\begin{figure}
    \includegraphics[width=0.85\textwidth]{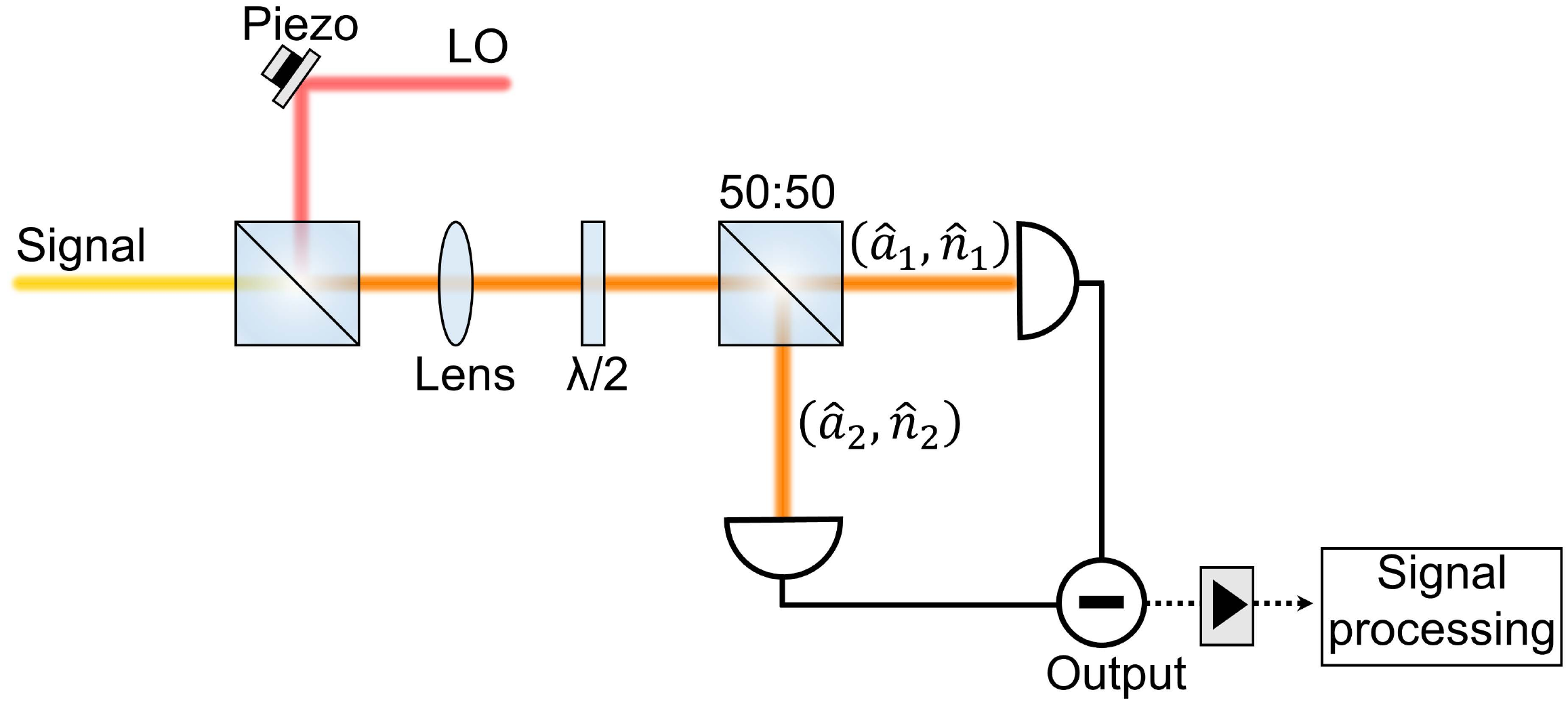}
    \caption{
        Schematic of a homodyne detection setup.
        The linearly polarized signal is combined with the orthogonally polarized local oscillator (LO) on a polarizing beam splitter. The orthogonal polarizations of the beams are rotated exactly to the diagonal basis, so that another polarizing beam splitter provides a 50:50 splitting ratio, before the signals are being detected by a balanced homodyne detector. $\hat{a}_{i}$ and $\hat{n}_{i}$ refer to the light fields and the photon numbers detected at each output port ($i=1,2$) of the homodyne detector.
        $\lambda$/2: half-wave plate.
    }\label{fig:Setup1Channel}
\end{figure}

    Homodyne detection is the ideal experimental tool to measure quadratures of the light field.
    It is phase-sensitive and enables quadrature measurements of very weak light fields combined with the possibility to single out individual modes of multi-mode light fields.
    A sketch of the experimental setup is shown in Fig. \ref{fig:Setup1Channel}.
    The signal of interest $\hat{a}_S$ is mixed with a strong, classical reference beam---the local oscillator (LO) $\hat{a}_{LO}$---on a 50:50 beam splitter.
    The light fields $\hat{a}_1$ and $\hat{a}_2$ at the two output ports of the beam splitter are given by
    \begin{equation}
    	\hat{a}_1=\frac{1}{\sqrt{2}}(\hat{a}_S+\hat{a}_{LO}), 
    	\text{ and }
    	\hat{a}_2=\frac{1}{\sqrt{2}}(\hat{a}_S-\hat{a}_{LO}).
    \end{equation}
    The photon numbers in the output ports are given by
    \begin{equation}
        \hat{n}_1=\hat{a}^\dagger_1 \hat{a}_1=\frac{1}{2}(\hat{a}_S^\dagger \hat{a}_S +\hat{a}_{LO}^\dagger \hat{a}_{LO}+\hat{a}_S^\dagger \hat{a}_{LO}+\hat{a}_{LO}^\dagger \hat{a}_S) 
    \end{equation}
    and 
    \begin{equation}
        \hat{n}_2=\hat{a}^\dagger_2 \hat{a}_2=\frac{1}{2}(\hat{a}_S^\dagger \hat{a}_S +\hat{a}_{LO}^\dagger \hat{a}_{LO}-\hat{a}_S^\dagger \hat{a}_{LO}-\hat{a}_{LO}^\dagger \hat{a}_S).
    \end{equation}
    These fields are detected by standard photodiodes.
    Additional electronics yield the difference photocurrent that is proportional to
    \begin{equation}
        \hat{n}_2-\hat{n}_1=\hat{a}^\dagger_S\hat{a}_{LO}+\hat{a}^\dagger_{LO}\hat{a}_S.
    \end{equation}
    The local oscillator is a strong classical coherent beam;
    so we can replace $\hat{a}_{LO}$ and $\hat{a}^\dagger_{LO}$ by their expectation values $\alpha_{LO}e^{\pm i\phi}$.
    The difference photocurrent then simplifies to
    \begin{equation}
        \hat{n}_2-\hat{n}_1=\alpha_{LO} (\hat{a}^\dagger_S e^{i\phi}+\hat{a}_S e^{-i\phi})=\sqrt{2}|\alpha|\hat{x}_\phi.
    \end{equation}
    
    As a consequence of the above considerations, homodyne detection measures the quadrature $\hat{x}_\phi$ of the signal mode, amplified by the amplitude of the local oscillator, $\sqrt{2}|\alpha|$.
    As the LO is an intense beam, this enormous amplification factor enables measurements of weak signal fields.
    The measurement is also phase-sensitive since the parameter $\phi$ is the relative phase between the signal and the LO which can be controlled, e.g., by shifting the LO phase using piezos.
    Finally, the measurement relies on the interference between the signal and the LO when passing through the beam splitter.
    The measurement is therefore mode sensitive.
    Also, only the part of the signal that overlaps spectrally, spatially, in polarization, temporally and in all other relevant degrees of freedom is amplified and detected.
    This renders homodyne detection particularly useful for spectroscopic applications.

    When considering conditional measurements again, it is important to note that the LO acts as a phase reference and all quadratures are defined with respect to this phase reference.
    As a consequence, the quadratures of light fields exactly in resonance with the LO do not rotate in phase space but remain stationary.
    Instead, the quadratures of all light fields not resonant with the LO will rotate with the difference frequency.
    As we show later, this turns out to be extremely helpful for studying dephasing processes.

\section{Continuous variable semiconductor spectroscopy}

    The experimental techniques outlined above have been utilized heavily in quantum optics to characterize light fields and their usefulness for certain tasks \cite{Breitenbach1997, Furusawa1998, Zhang2003, Weedbrook2004, Appel2008,Masada2015,Albarelli2018,Yadin2018}.
    Although the strong mode selectivity that is enforced by the local oscillator and the high temporal resolution that can be achieved are ideal features for semiconductor spectroscopy, homodyne detection has found far less widespread applications in this field.
    In this section, we discuss the different requirements for applying homodyne detection in quantum optics and semiconductor spectroscopy and explore how recent technological progress significantly broadens the scope of homodyne detection in spectroscopy.

    Probably the most significant difference between quantum optics and semiconductor spectroscopy is the subject of interest.
    In many scenarios in quantum optics, researchers are interested in the carefully prepared state of the light field itself.
    Therefore, it is paramount to tailor the temporal and modal overlap between the local oscillator and the light field of interest to optimize the detection efficiency, which is crucial for identifying quantum features of the field.
    In semiconductor spectroscopy, the light field is rather an ancilla or messenger that carries information about the spontaneous dynamics of a system of interest which is monitored passively.
    The dynamics of a semiconductor system with a sophisticated level structure is mapped onto an inherently multi-mode optical signal, showing ultrafast dynamics which is usually the quantity of interest in semiconductor spectroscopy.

    The passive monitoring approach employed in semiconductor spectroscopy introduces two important differences to quantum optics.
    First, the dynamics of interest are usually spontaneous; that is, there is no fixed phase between any external local oscillator and the light fields emitted by the semiconductor sample.
    Any set of single-channel quadrature measurements that has been accumulated over a period longer than the coherence time of the signal is therefore necessarily phase-averaged.
    Second, in quantum optics, temporal resolution typically refers to the duration $t_{LO}$ of the local oscillator.
    A faithful quadrature measurement requires that the signal light field does not change significantly within $t_{LO}$.
    In semiconductor spectroscopy, the dynamics of the system interacting with the light field are of interest.
    Identifying these dynamics requires more than one measurement.
    The change in the light field between the two measurements then reflects the change in the system itself.
    The effective temporal resolution of the system is now determined by two factors.
    Again, the system should not change significantly during the local oscillator duration $t_{LO}$ to ensure that a single measurement of the signal light field corresponds to a quasi-static snapshot of the studied system.
    However, now the minimal obtainable delay $\tau_{LO}$ between two consecutive measurements determines the temporal resolution at which the dynamics of the system can actually be monitored.

    Multimode detection utilizing several homodyne detection channels is essential for tackling these two challenges.
    When distributing a semiconductor signal across several homodyne detection channels, the signal does not have any fixed phase relative to the LOs, but the different LO beams share a precisely adjustable relative phase.
    If a set of homodyne measurements is used to estimate the relative phase between the signal and the LOs, all consecutive measurements can be performed with respect to this ad hoc phase reference.
    In this manner, information about the signal phase indeed becomes available.
    Further, delay lines may be utilized to interrogate the signal at variable delays $\tau_{LO}$,  easily reaching the sub-picosecond scale.
    By changing the modes of the different LOs, it also becomes possible to study correlations between different states and emission channels of the semiconductor system under study.
    We would like to emphasize that the frequently encountered multimode homodyne detection schemes in quantum optics (see, e.g., \cite{Takase2019, Morin2013}) are exciting experiments but unfortunately not useful at all for multimode semiconductor spectroscopy as they do not offer any relevant temporal resolution in terms of $\tau_{LO}$. 
    In such experiments, $\tau_{LO}$ is usually on the order of tens of nanoseconds, which is two to five orders of magnitude too slow compared to the relevant timescales in semiconductors.
    Indeed, the initial groundbreaking work on ultrafast multimode homodyne detection was already performed in the 1990s \cite{Raymer1996, McAlister1997}.
    Despite these impressive demonstrations, ultrafast multimode homodyne spectroscopy has not been developed further in quantum optics.
    Significant extensions have been realized in semiconductor physics more than 15 years later \cite{Roumpos2013}.
    Currently, renewed interest in ultrafast homodyne spectroscopy is arising that is fueled by the interest in continuous-variable effects, such as squeezing in semiconductor systems \cite{Fox1995,Boulier2014,Denning2022}, and also by the arrival of ultrafast analog-to-digital converters that can cope with huge data rates beyond 5\,GB per second.
    At these data rates, full time-tagged sets of individual multi-channel quadrature measurements may be recorded instead of just histograms.
    In this way, all correlations between measurements are conserved, which opens up the path to the postselective measurements as described in this review.
    Combined with on-chip data analysis, it will also become possible to employ ultrafast homodyne detection as a real-time spectroscopy method, which is of fundamental importance for real-world applications.

    In the following, we summarize several applications of ultrafast homodyne spectroscopy in order of increasing complexity.
    We review its real-time capabilities in single-channel ultrafast correlation measurements, we outline how homodyne detection can be used to apply quantum resource theory to semiconductors, and, 
    finally, we discuss a full time-resolved multi-channel experiment that enables postselective phase-resolved spectroscopy.

\section{Real-time monitoring of intensity correlations}

    Photon statistics are essential tools for characterizing technologically relevant quantum light fields.
    The quantity studied most frequently in experiments is the second-order correlation function 
    \begin{equation}
        \label{eq:longg2}
        g^{(2)}(\tau)=\frac{\langle \hat{a}^\dagger(t) \hat{a}^\dagger(t+\tau) \hat{a}(t+\tau)\hat{a} \rangle }{\langle \hat{a}^\dagger (t) \hat{a}(t)\rangle \langle \hat{a}^\dagger(t+\tau) \hat{a}(t+\tau) \rangle},
    \end{equation}
    where the operators describe a light field mode of interest.
    $g^{(2)}(\tau)$ characterizes photon number fluctuations and can be interpreted as the conditional relative probability for detecting a photon at time $\tau$ given that another photon was already detected at time $t=0$.
    Especially the zero-delay, second-order intensity correlation function $g^{(2)}(0)$ is studied on a regular basis as values of $g^{(2)}(0)$ corresponding to 2, 1 and 0 are good indicators for thermal, coherent and single-photon states, respectively.
    It is also the ideal tool to study the onset of coherent emission across the threshold in lasing devices, especially for highly efficient nanoscale devices \cite{Assmann2010,Chow2014}.

    The averaging in Eq. (\ref{eq:longg2}) commonly runs over all available times $t$.
    This approach pertains to the assumption that the light field of interest itself is long-term stable in the sense that its mean photon number and its noise properties do not change with time.
    When considering realistic devices, especially nanolasers, this is not necessarily true.
    Feedback and spontaneously arising perturbations can have a significant detrimental effect on the performance of the emitter.
    The fast fluctuations characterized by $g^{(2)}(0)$ themselves are then subject to additional fluctuations on slower timescales.
    One may gain experimental access to these noise dynamics by averaging over time windows of fixed length $\Delta t$ in Eq. (\ref{eq:longg2}) instead,
    \begin{equation}
        \label{eq:shortg20}
        g^{(2)}(0)=\frac{\langle \hat{a}^\dagger \hat{a}^\dagger \hat{a}\hat{a} \rangle_{\Delta t} }{\langle \hat{a}^\dagger \hat{a}\rangle_{\Delta t}^2}.
    \end{equation}

    Unfortunately, in the most commonly encountered DV $g^{(2)}$-measurement setting, using a Hanbury Brown-Twiss (HBT) type setup \cite{Brown1956}, detector deadtimes and the avoidance of pile-up effects require quite long integration times to obtain a single measurement point so that the $\Delta t$ that may be obtained this way is typically on the order of minutes or at least seconds.

    Interestingly, the expectation values of the operator combinations appearing in Eq. (\ref{eq:shortg20}) can also be determined using a CV approach.
    When performing homodyne detection without any fixed phase relationship between the local oscillator and the signal so that all relative phases are sampled equally, the higher-order moments of the measured set of quadratures $\tilde{x}$ are directly proportional to the sought-after expectation values \cite{McAlister1997,Roumpos2013,Lueders2018}.
    In particular, one obtains
    \begin{equation}
        \label{eq:QuadN}
        \langle \hat{a}^\dagger \hat{a}\rangle=\frac{1}{2}(2\langle \tilde{x}^2 \rangle-1)
    \end{equation}
    and
    \begin{equation}
        \label{eq:QuadVar}
        \langle \hat{a}^\dagger \hat{a}^\dagger \hat{a} \hat{a}\rangle=\frac{2}{3}\langle \tilde{x}^4 \rangle-2 \langle \hat{a}^\dagger \hat{a}\rangle-\frac{1}{2}.
    \end{equation}
    The homodyne measurement reduces the $\Delta t$ that can be reached by several orders of magnitude compared to experiments based on photon counting.
    It suffices to average over about $10^{3}$ to $10^{4}$ quadratures to obtain a reliable $g^{(2)}(0)$-value.
    Considering typical laser repetition rates of about 80\,MHz, this results in $\Delta t$ as low as 10-100\,$\mu$s.
    The prerequisite that all relative phases between the local oscillator and the signal must be sampled equally is automatically fulfilled as long as the signal coherence time is much shorter than $\Delta t$.

    To demonstrate the capabilities of homodyne correlation measurements, we used it to investigate the emission of an 830\,nm cw external-cavity diode laser continuously.

    First, we perform a standard analysis of the onset of lasing by using all of the available measurement values to determine the mean photon number and the intensity correlation of the light field emitted by the diode laser.
    Both exhibit a typical laser threshold at a current of about 70.5\,mA, as indicated by a strong increase in the emitted photon number and a steep drop in $g^{(2)}(0)$ from 2 to 1.

\begin{figure}
    \includegraphics[width=0.95\textwidth]{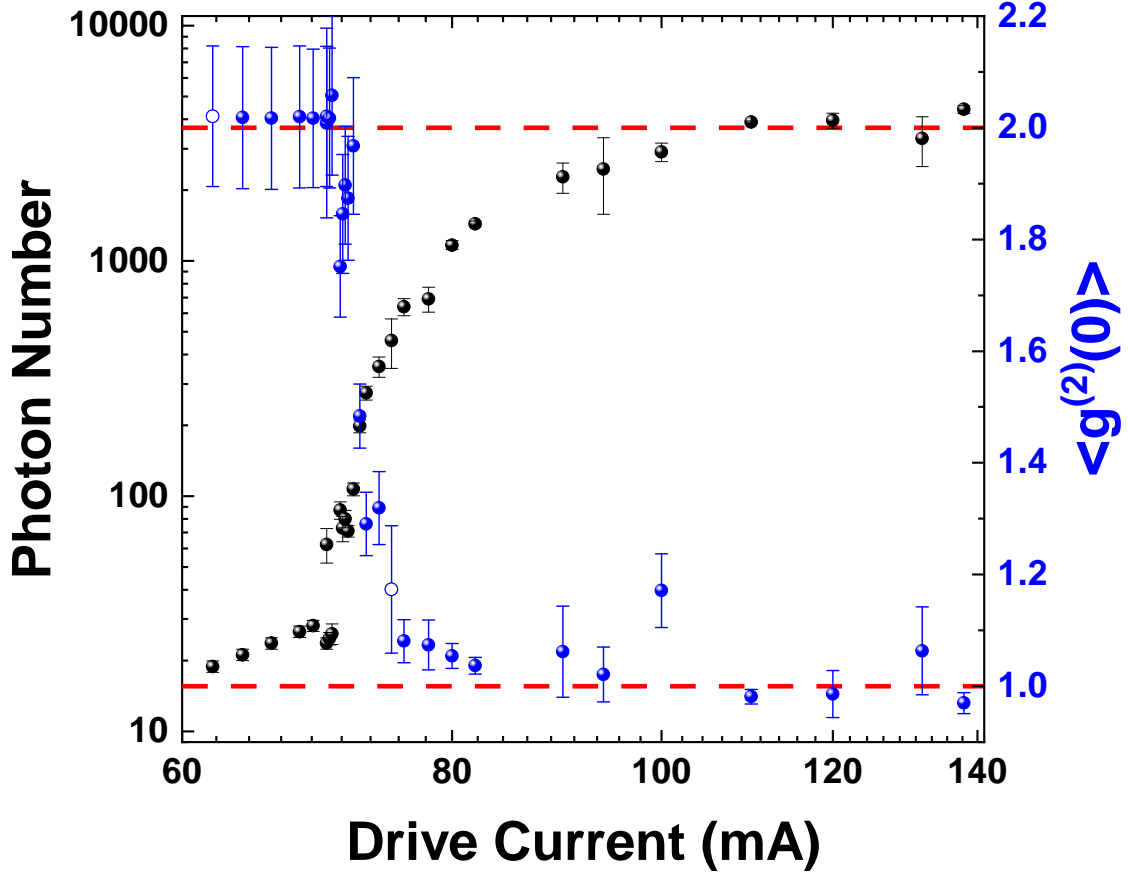}
    \caption{
        Emitted photon number (black dots) and second-order intensity correlation (blue dots) for several drive currents of a cw external-cavity diode laser.
        Red dashed lines mark the expected $g^{(2)}(0)$-values for coherent and thermal emission.
        The lasing threshold arises at 70.5\,mA.
    }\label{fig:IOCurve}
\end{figure}

\begin{figure}
    \includegraphics[width=0.75\textwidth]{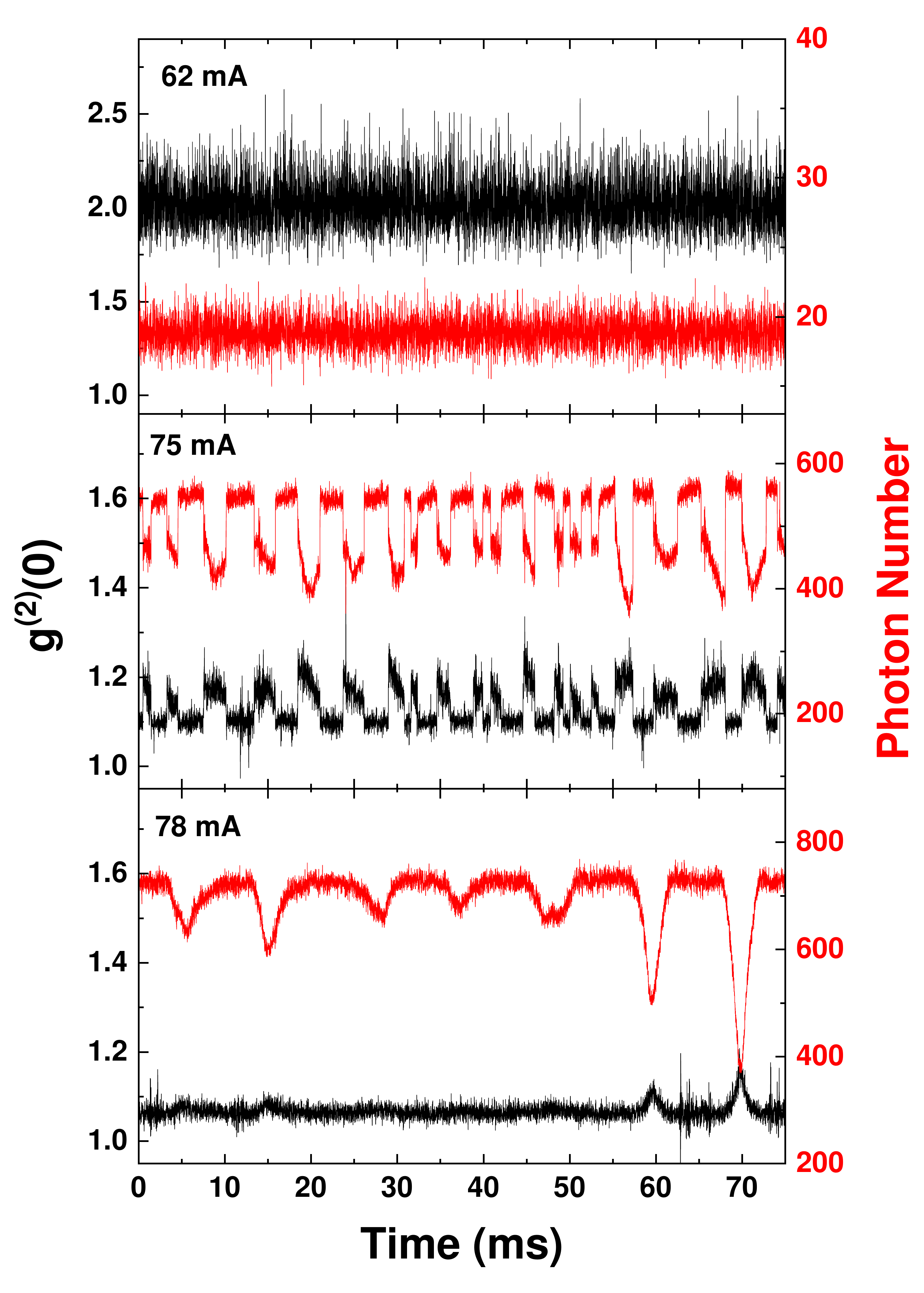}
    \caption{
        Emitted photon numbers (red lines) and $g^{(2)}(0)$ (black lines) for driving the laser below threshold, slightly above threshold and above threshold.
    }\label{fig:TimeSeries}
\end{figure}

    Next, we divide the whole time series into discrete segments of lengths of 10\,$\mu$s each and determine both the mean photon numbers and $g^{(2)}(0)$ for each segment separately.
    The results of this approach are shown in Fig. \ref{fig:TimeSeries} for several characteristic drive currents.

    Below threshold, at a drive current of 62\,mA, the photon number is constantly moderately low at about 20 photons, and $g^{(2)}(0)$ shows a constant value of 2, indicating thermal emission.
    The additional benefit of investigating a series of segments becomes apparent for a drive current of 75\,mA, which is in the middle of the threshold region.
    Here, the emission becomes unstable and the emitted photon number shows almost periodic jumps between 400 and 600 photons.
    Interestingly, $g^{(2)}(0)$ undergoes similar jumps which are exactly correlated with the steps in the photon number.
    The intensity noise is reduced when the photon number becomes increased and vice versa.
    This is a typical indicator for gain competition between different resonator modes or polarizations.
    Finally, farther above threshold at a drive current of 78\,mW, the photon number stabilizes at a value slightly above 700 and $g^{(2)}(0)$ stays at a level of about 1.06.
    Drops in the photon number still occur but take place at a significantly lower rate compared to operating the laser directly in the threshold region.

    When evaluating the performance of a laser device, the time-resolved measurements of the intensity correlation made possible by homodyne detection are clearly superior to standard HBT-like DV approaches.
    CV spectroscopy offers the possibility to access experimental parameters, such as the mean jump rates between different emission levels, a distribution of the mean dwelling times and other relevant parameters.
    It also makes it much easier to observe rare noise-driven perturbations, which often go unnoticed in HBT-type experiments, but are detrimental when operating the laser and may even indicate the onset of damage.
    In this way, it also becomes possible to unravel intrinsic noise of the device and noise arising due to external influences as these typically occur on different timescales \cite{Lueders2020}. 

\section{Tailored phase-space distributions for quantum resource theories of semiconductors}

    In the last section, we discussed the performance of a laser diode in terms of its equal-time second-order intensity correlation.
    The latter is considered as an unambiguous identifier for distinguishing lasing from thermal emission in the community working on highly efficient lasers.
    Along similar lines, a value of $g^{(2)}(0)$=0 is a good identifier for an ideal single-photon state.
    It is also firmly established that how closely a real single-photon source approaches the ideal value of 0 provides a quantitative measure for how well that single-photon source performs in quantum technology tasks thar require single photons.
    It is therefore considered both a quantifier for the non-classicality of a light field and its resourcefulness in tasks that rely on single photons.

    Besides Fock states, also other non-classical light fields, such as entangled photon states and squeezed light fields, qualify as valuable resources in quantum technology tasks.
    Entangled photons, for example, enable quantum teleportation tasks, and squeezed light enhances phase resolution.
    With the emergence of quantum technologies, it has become clear that it makes sense to define the resourcefulness of a state---and therefore also its non-classicality---with respect to its usefulness in a certain quantum information task.
    Finding such quantitative notions of usefulness is the domain of quantum resource theory.

    It ought to be emphasized that this approach results in terminology that differs from the one commonly utilized in quantum optics.
    In quantum optics, photon-number states are considered non-classical while superpositions of Fock states forming coherent states are considered as the most classical light fields.
    In contrast, for matter-based systems, such as atoms, a single atom in a well-defined stable state is deemed classical while superpositions thereof are non-classical and useful in quantum information tasks.
    For example, it is possible to directly convert the quantum coherence carried by such a superposition state into entanglement \cite{Killoran2016}.
    Still, there are a manifold of applications in quantum technologies where old-fashioned notions of classicality are unhelpful.
    For example, in quantum random number generation \cite{Symul2011,Ferreira2016} and quantum cryptography \cite{Pirandola2020} with coherent light fields, deemed classical in quantum optics, coherent states are the fundamental quantum resources in these realizations.
    In practice, most quantum technology applications will likely be hybrid quantum systems, which consist of at least two distinct physical platforms that are able to carry, process and transmit the same kind of quantum information content.
    Light-matter interfaces are prototypical examples of such hybrid systems.
    Considering the mutually exclusive notions of non-classicality for light fields and matter particles outlined above, it becomes clear that an agnostic approach that can be tailored to a certain application utilizing quantum technologies is preferable.
    The concept of quantum coherence which allows for general classical references achieves exactly that goal.
    In the following, we investigate the quantum coherence of exciton-polaritons, nontrivial semiconductor quasiparticles consisting of cavity photons and excitons, and thus archetypal hybrid systems.

\subsection{Quantum-information-based quantum coherence}\label{QCDefinition}


	Unlike in the context of quantum optics, the notion of quantum coherence is based on superposition of computational basis states in quantum information science \cite{Streltsov2017,Chitambar2019}.
	For quantized fields, the excitations of that field $|n\rangle$ can be considered as such computational basis states, such as applied in Ref. \cite{Lueders2021}.
	Physically speaking, this indicates superpositions in the particle picture $|n\rangle$.
	For example, the state $(|0\rangle+|1\rangle)/\sqrt2$ represents a qubit state that, in contrast to classical bit values, is in a superposition of true ($1$) and false ($0$).

	Extending the concept of superpositions to mixed states, we find that states that are a classical statistical mixture of computational basis states are free of superpositions, defining incoherent states \cite{Baumgratz2014,Levi2014,Sperling2015,Winter2016},
	\begin{equation}
		\hat\rho_\mathrm{inc.}=\sum_{n}p_n|n\rangle\langle n|,
	\end{equation}
	with $p_n\geq0$ for all $n$ and $\sum_n p_n=1$.
	A state that cannot be diagonalized in this form requires quantum superpositions and, thus, carries a certain amount of quantum coherence.
	An example is $\hat\rho=(|0\rangle\langle 0|+|1\rangle\langle 1|+q|1\rangle\langle 0|+q^\ast|0\rangle\langle 1|)/2\neq\hat\rho_\mathrm{inc.}$ for $q\in\mathbb C$ with $0<|q|\leq 1$.

	To quantify the amount of quantum coherence, a variety of distance measures of a given state from the set of incoherent states can be introduced \cite{Baumgratz2014,Sperling2015}.
	One example is given in terms of the Hilbert-Schmidt distance,
	\begin{equation}
		\mathcal C(\hat\rho)
		=\sup_{\hat\rho_\mathrm{rec.}}\left\{\mathrm{tr}\left(\left[\hat\rho-\hat\rho_\mathrm{inc}\right]^2\right)\right\}
		=\mathrm{tr}\left(\hat\rho^2\right)
		-\mathrm{tr}\left(\left[\sum_np_n|n\rangle\langle n|\right]^2\right)
		=\sum_{m,n:m\neq n}|\rho_{m,n}|^2,
	\end{equation}
	with $\hat\rho=\sum_{m,n}\rho_{m,n}|m\rangle\langle n|$, where $\rho_{m,n}=\rho_{n,m}^\ast$, and where the minimal distance is achieved for $p_n=\langle n|\hat\rho|n\rangle=\rho_{n,n}$.
	Evidently, $\mathcal C$ is zero if and only if all off-diagonal elements vanish---meaning the state is incoherent---and is nonzero otherwise, requiring superpositions to express the (mixed) quantum state under study.

	The notion of quantum coherence straightforwardly generalizes to multipartite systems \cite{Sperling2015}.
	That is, for $N$ quantized fields, the statistical mixture of number states marks an incoherent ensemble,
	\begin{equation}
		\hat\rho_\mathrm{inc.}=\sum_{n_1,\ldots,n_N} p_{(n_1,\ldots,n_N)}|n_1,\ldots,n_N\rangle\langle n_1,\ldots,n_N|,
	\end{equation}
	And the contributions of off-diagonal density matrix entries quantifies the multipartite quantum coherence,
	\begin{equation}
        \label{eq:CohNorm}
		\mathcal C(\hat\rho)=\sum_{\substack{
			n_1,\ldots,n_N,m_1,\ldots,m_N:
			\\
			(n_1,\ldots,n_N)\neq(m_1,\ldots,m_N)
		}}|\rho_{(n_1,\ldots,n_N),(m_1,\ldots,m_N)}|^2.
	\end{equation}
	In the context of multipartite quantum coherence, it is noteworthy that seminal quantum correlations as represented by entanglement are, in fact, a subclass of quantum coherence phenomena \cite{Killoran2016,Chitambar2016,Qiao2018,Ma2016}.
	That is, the Bell state $(|0,0\rangle+|1,1\rangle)/\sqrt{2}$ is a quantum superposition of computational basis states, thus exhibiting useful quantum coherence quantum quantum information processing.
	At the same time, $(|0,0\rangle+|0,1\rangle)/\sqrt2=|0\rangle(|0\rangle+|1\rangle)/\sqrt2$ also exhibits quantum coherence but not entanglement, while being resourceful for quantum processing, still.
	In general, every entangled state exhibits quantum coherence but not necessarily the other way around.

	Finally, we can exemplify the impact of a typical decoherence process on the amount of quantum coherence.
	This is exemplified by the pure dephasing channel
	\begin{equation}
		\Lambda_\sigma(\hat\rho)
		=\int_{-\sigma/2}^{\sigma/2}\frac{d\varphi}{\sigma}\,e^{i\varphi\hat n}\hat\rho e^{-i\varphi\hat n}
		=\sum_{m,n}\mathrm{sinc}\left(\sigma\frac{m-n}{2}\right)\rho_{m,n}|m\rangle\langle n|,
	\end{equation}
	where $\hat n$ is the particle-number operator, $\sigma$ is the width of the phase interval over which we average, and $\mathrm{sinc}(z)=\sin(z)/z$ denotes the cardinal sine function, with $\mathrm{sinc}(0)=1$.
	For $\sigma=0$, no decoherence occurs, $\mathcal C(\Lambda_0(\hat\rho))=\mathcal C(\hat\rho)$; for $\sigma=2\pi$, full decoherence leads to an incoherent channel output for all inputs, $\mathcal C(\Lambda_{2\pi}(\hat\rho))=0$.
	Since $|\mathrm{sinc}(z)|^2<1$ for $z\neq0$, we generally find a decreased output quantum coherence, $\mathcal C(\Lambda_\sigma(\hat\rho))<\mathcal C(\hat\rho)$ for $\sigma>0$.

	Furthermore, we may use common phase-space representations to quantify the quantum coherence.
	For instance via the Glauber--Sudarshan distribution $P(\alpha)$ \cite{Glauber1963,Sudarshan1963}, allowing us to expand the state as
	\begin{equation}
        \label{eq:GS}
		\hat\rho=\int d^2\alpha\,P(\alpha) |\alpha\rangle\langle\alpha|,
	\end{equation}
	using coherent states $|\alpha\rangle=\sum_{n}e^{-|\alpha|^2/2}\alpha^n/\sqrt{n!}|n\rangle$ with a coherent amplitude $\alpha\in\mathbb C$.
	In particular, incoherent states, being diagonal in the number basis by definition, are rotationally invariant, $P(\alpha)=P(|\alpha|)$.
	Any structure in the phase $\phi=\arg(\alpha)$ that deviates from a uniform distribution is thus a signature of quantum coherence in phase space.
	The circular variance,
	\begin{equation}
		\mathrm{Var}(\phi)=1-\left|
			\int d^2\alpha\,P(\alpha)\frac{\alpha}{|\alpha|}
		\right|
	\end{equation}
	is a operational means to quantify the quantum coherence through phase-space distributions \cite{Lueders2023}.
	The larger the value of the circular variance, saturating at a value of one for incoherent states, the lower the quantum coherence.

	In summary, quantum coherence quantifies the amount of quantum superpositions of computational basis states.
	Since classical computation is limited to individual computational states and statistical ensembles thereof, while explicitly disallowing superpositions, quantum coherence assesses the overhead of resources provided by quantum physics.
	Entanglement is one manifestation of quantum coherence in multipartite systems.
	Dephasing reduces quantum coherence by suppressing off-diagonal entries in the density matrix in the computational basis, here chosen as the number basis of one or multiple quantized field in Fock space.
	Common phase-space representations of quantum states encode quantum-coherence features in terms of the non-uniformity in their phase distribution.
	Both such phase-dependent phase-space structures and the off-diagonal quantum-superposition terms in the statistical operator are thus complementary, yet equally intuitive means to explore the physics of quantum superpositions together with the quantum superposition's operational usefulness as a resources for quantum information science and technology.

\subsection{Phase-space distributions}

    So far, we mostly used DV descriptions of the light field in terms of photon numbers, such as $g^{(2)}$ and the density matrix in the Fock basis, but we utilized CV, such as field quadratures, to calculate them.
    Next, we will introduce CV representations of light fields in addition to the aformentioned Glauber-Sudarshan distribution, such as the Wigner and Husimi-Q distributions.
    Recall that their information content is exactly equivalent to the density matrix in the Fock basis.
    A natural basis in the CV regime are coherent states $|\alpha\rangle$, exhibiting minimal uncertainty for all pairs of orthogonal quadratures, $\hat{x}_\phi$ and $\hat{x}_{\phi+\pi/2}$, which are commonly termed $\hat{q}$ and $\hat{p}$.
    Note that the coherent states are not orthogonal and therefore form an overcomplete generating set.
    This property is at the heart of several concepts in quantum cryptography \cite{Gabriel2010,Avesani2018,Qi2017,Thewes2019a}.
    Based on coherent states, the Glauber--Sudarshan $P$ distribution in Eq. (\ref{eq:GS}) is already a good tool to identify non-classical states in quantum optics.
    Any state which yields a $P$ distribution containing non-negative values is considered non-classical in quantum optics.
    However, the $P$ distribution may become highly singular\cite{Sperling2016} and is not directly accessible experimentally.

    Smoothing the $P$ distribution by convoluting it with a Gaussian kernel removes singularities and yields the Wigner function $W(\alpha)$ \cite{Wigner1932}.
    Negativities in the Wigner function serve as an sufficient criterion for light fields considered non-classical in quantum optics.
    Also, it is partially accessible in experiments.
    When performing a set of homodyne detection quadrature measurements with a fixed phase between the local oscillator and the signal, the resulting quadrature distribution corresponds to one of the marginals of the Wigner function.
    It is the projection of the two-dimensional Wigner function onto the axis defined by the relative phase between the signal and the local oscillator.
    One can record sets of such projected quadrature distributions for varying relative phases between local oscillator and signal and thereby reconstructs the full two-dimensional Wigner function by means of quantum state tomography using maximum-likelihood methods \cite{Lvovsky2004} and pattern functions \cite{PhysRevA.52.4899}.

    When the relative phase between the local oscillator and the signal is not fixed or known, which is usually the case in semiconductor spectroscopy, it is more common to measure the Husimi $Q$ function instead \cite{Husimi1940}.
    Formally, it is given by convoluting the Wigner function again with a Gaussian kernel,
    \begin{equation}
        Q(\alpha)=\int  d^2 \beta \frac{2}{\pi} W(\beta)e^{-2|\alpha-\beta|^2}=\frac{\langle\alpha|\hat{\rho}|\alpha\rangle}{\pi},
    \end{equation}
    where the latter expression corresponds effectively to evaluating the density matrix on the basis of coherent states.
    From an experimental point of view, the $Q$ function corresponds to the two-dimensional probability distribution for simultaneously measuring a certain combination of orthogonal quadratures $\hat{x}$ and $\hat{p}$, when splitting a signal of interest into two beams and performing homodyne detection with two local oscillators whose phases are exactly orthogonal to each other.
    Splitting the signal light field at a beam splitter necessarily includes mixing the state with the vacuum state that is present at the other input port of the beam splitter.
    This increases the uncertainties of both quadratures in full analogy to a joint measurement of position and momentum performed on a harmonic oscillator.
    As a consequence, $Q$ is a non-negative distribution.
    Accordingly, due to the absence of negativities, identifying non-classical states in quantum optics becomes more difficult based on $Q(\alpha)$, as compared to $W(\alpha)$.
    However, the frequently encountered claim that $Q(\alpha)$ cannot be used to identify non-classical states is not tenable.
    Constructing suitable identifiers of non-classicality for $Q$ simply requires more care.
    Regardless, the information contained in $P$, $W$, $Q$ and $\hat{\rho}$ is identical.

    In the absence of a fixed phase between the local oscillator and the signal, the relationship between phase-randomized single-quadrature measurements and the Wigner function is highly nontrivial, but a set of phase-randomized pairs of quadratures obtained, where the phases of the two local oscillators are kept strictly orthogonal, will yield the phase-averaged Husimi $Q$ function, which can be calculated analytically easily, too.

\subsection{Quantum coherence of a polariton condensate}
    
    In the following, we determine the quantum coherence of a prototypical hybrid system: an exciton-polariton condensate.
    The planar microcavity we use features two DBR mirrors made of alternating layers of $\mathrm{Al}_{0.2}\mathrm{Ga}_{0.8}\mathrm{As}$ and AlAs and shows a quality factor on the order of 20000.
    Four GaAs quantum wells are contained in the central $\lambda/2$ cavity enclosed by the mirrors.
    The Rabi splitting of the sample amounts to 9.5\,meV, and the exciton-cavity detuning is about -6.4\,meV.
    Detailed information about the sample can be found in \cite{Ma2020}.
    We operate the sample at 10\,K and excite the sample non-resonantly using a continuous-wave (cw) laser, so that all coherence in the system must build up spontaneously and is not inherited from the pump beam \cite{Keeling2007,Kasprzak2006}.
    We use a Gaussian spot with a diameter of 70\,$\mu$m to excite the sample and filter the emission in momentum space so that only the emission around $k=0$ is investigated. 

\begin{figure}
    \includegraphics[width=0.75\textwidth]{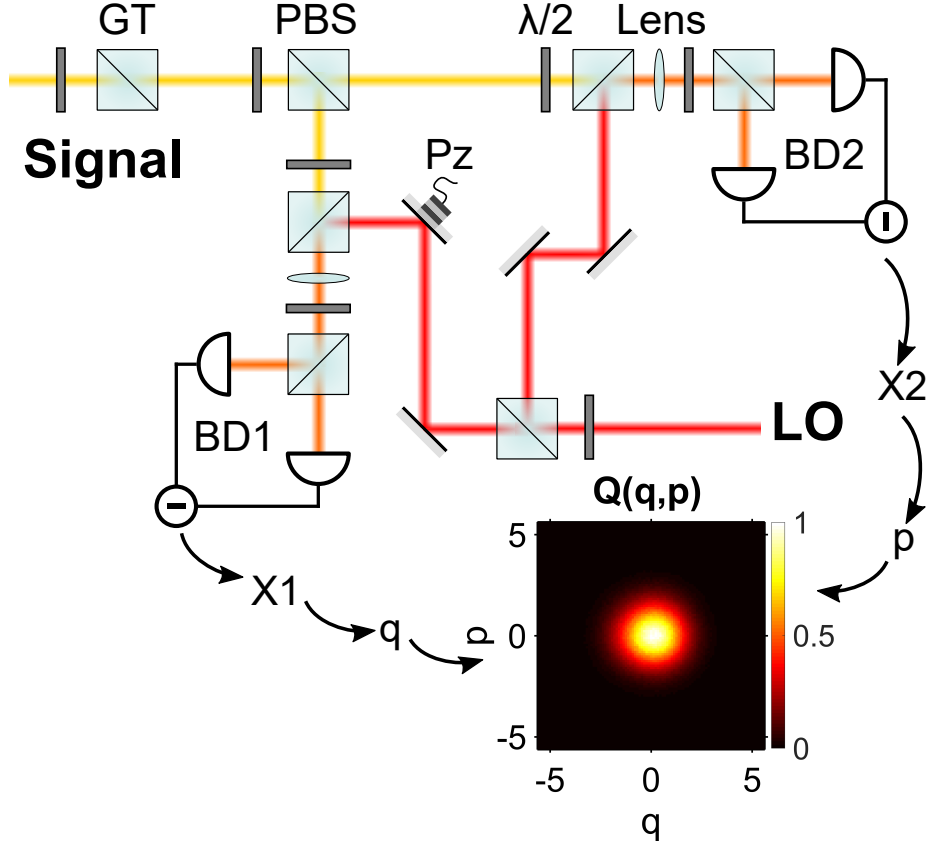}
    \caption{
        Two-channel homodyne detection setup.
        The polariton signal is split into two channels.
        Each channel is interfered with the local oscillator (LO) and detected at a balanced homodyne detector (BD).
        The path of the LO in channel one contains a piezo mirror (Pz) for setting the relative phase between the two LOs.
        The two channels measure quadratures $X_1$ and $X_2$.
        They are filtered in postselection to ensure that only quadrature pairs $\hat{q}$ and $\hat{p}$ for fully orthogonal LO settings are considered.
        The quadrature pairs are then binned to a histogram $Q(q,p)$, which corresponds to the Husimi $Q$ function when normalized.
        (PBS: polarizing beam splitter; GT: Glan-Thompson prism; $\lambda$/2: half-wave plate). Reproduced from \cite{Lueders2021}.
    }\label{fig:Setup2Channel}
\end{figure}

    We use an eight-port homodyne detection setup \cite{Qi2020} (see Fig. \ref{fig:Setup2Channel}), where the emission is split into two beams, and we perform homodyne detection on the ground state emission with two local oscillators of orthogonal phase, as outlined above.
    As a result, we obtain the phase-averaged Husimi $Q$ function, which already contains sufficient information to reconstruct the amount of quantum coherence present in the system.

\begin{figure}
    \includegraphics[width=0.95\textwidth]{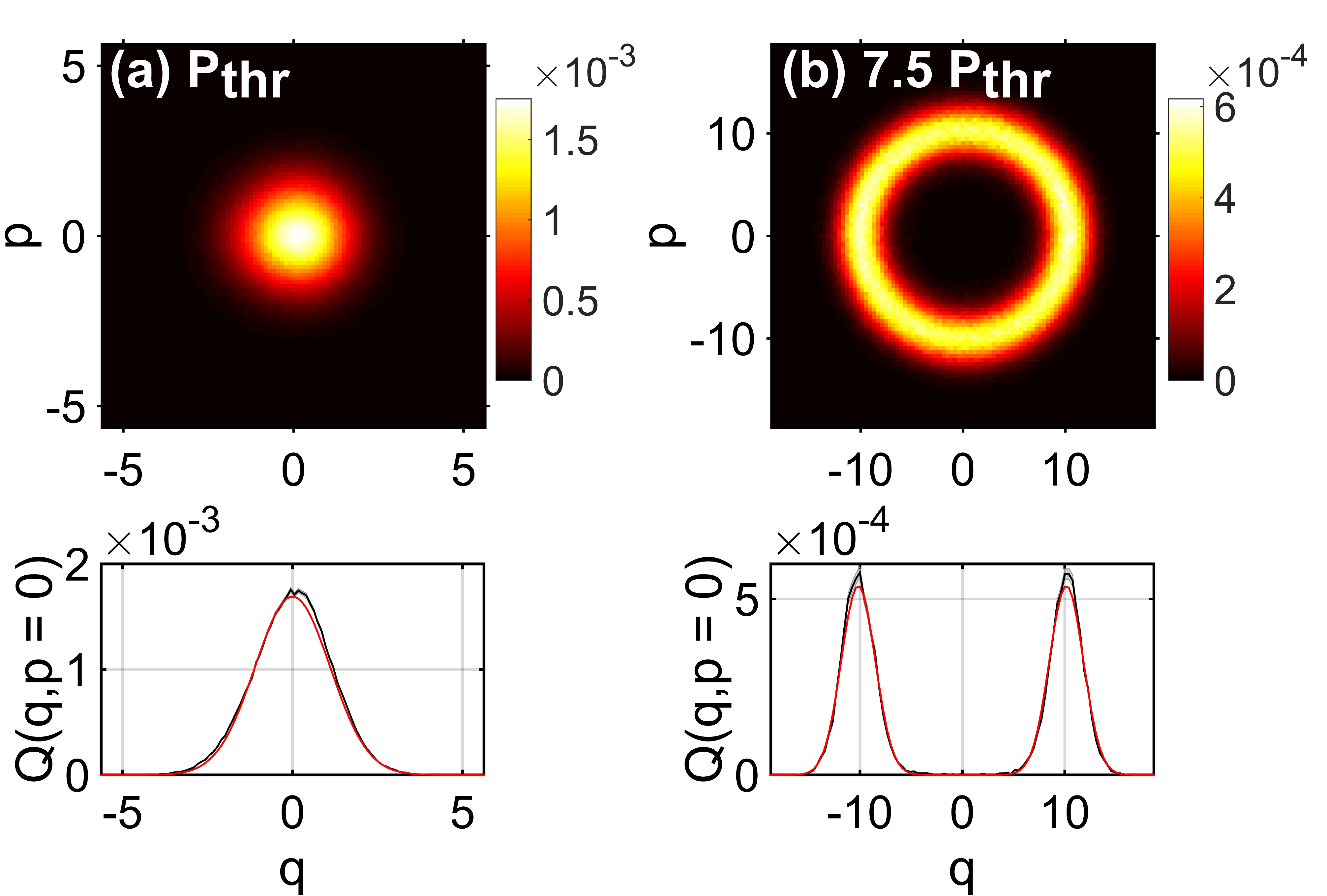}
    \caption{
        Measured phase-averaged Husimi functions for (a) $P=P_{\mathrm{thr}}$ and (b) $P=7.5\,P_{\mathrm{thr}}$.
        The bottom plots show cuts along the $q$-axis for $p=0$.
        Black lines represent the measured data, while red lines show the fit function for a displaced thermal state.Reproduced from \cite{Lueders2021}.
    }\label{fig:Husimis-PhaseAveraged}
\end{figure}

    An example of two measured phase-averaged Husimi functions for different pump powers is shown in Fig. \ref{fig:Husimis-PhaseAveraged}.
    At the threshold power of $P_{\mathrm{thr}}=30\,$mW, the resulting distribution is a Gaussian distribution centered at the origin of phase space.
    In contrast, at 7.5\,$P_{\mathrm{thr}}$, a displacement becomes apparent and the distribution takes the shape of a ring with a certain width.
    We will now show that the mean values and variances of the quadratures contain sufficient information to model the quantum coherence available in the state. 

\section{Quantum coherence of displaced thermal states}\label{SteadyQC}

    Although polariton condensation relies on stimulated scattering instead of stimulated emission, from a quantum-optical point of view, in many cases the emission can be characterized in a manner similar to laser light.
    We can identify the ground state of the polariton condensate with a single bosonic mode.
    Below threshold, the emission is dominated by incoherent spontaneous thermal emission.
    Above threshold, stimulated scattering becomes dominant, and a coherent population of polaritons builds up.
    The relative importance of both contributions depends on the exact pump power used.
    This kind of emission can be described by a displaced thermal state, which has a Glauber-Sudarshan distribution given by
    \begin{equation}
        P_{\mathrm{dt}}(\alpha)=\frac{\exp{(-(|\alpha-\alpha_0|^2/\bar{n}))}}{\pi \bar{n}},
    \end{equation}
    where $\alpha_0$ corresponds to the coherent amplitude and $\bar{n}$ corresponds to the thermal photon number that defines the width of the distribution.
    The density matrix of a displaced thermal state $\hat{\rho}_{\mathrm{dt}}$ can be obtained via Eq. (\ref{eq:GS}).
    In the limit of thermal emission for pumping below threshold, we expect the coherent displacement to vanish, so $\alpha_0\rightarrow 0$.
    This results in a thermal state which shows no coherences,
    \begin{equation}
        \hat{\rho}_{\mathrm{th}}=\frac{1}{\bar{n}+1} \sum_n \left(\frac{\bar{n}}{(\bar{n}+1)}\right)^n |n\rangle\langle n |.
    \end{equation}
    If, instead, coherence dominates and the thermal photon number becomes negligible, $\bar{n}\rightarrow 0$, one approaches the coherent state
    \begin{equation}
        \hat{\rho}_{\mathrm{coh}}=|\alpha_0\rangle \langle \alpha_0|,
    \end{equation}
    which shows off-diagonal elements of the form $
    \rho_{m,n}=\frac{e^{-|\alpha_0|^2} \alpha_0^m \alpha_0^{*n}}{\sqrt{m! n!}}
    $ and a coherent photon number $|\alpha_0|^2$.
    In fact, this state shows a substantial amount of quantum coherence.
    In order to assess this, we need to determine both the coherent and the thermal photon number of the displaced state.

    Moreover, the density matrix of the displaced thermal state outlined above directly yields the Husimi function,
    \begin{equation}
        Q_{\mathrm{dt}}(\alpha)=\frac{\langle\alpha|\hat{\rho}_{\mathrm{dt}}|\alpha\rangle}{\pi}=\frac{\exp{(-[(|\alpha-\alpha_0|^2)/(\bar{n}+1)])}}{\pi(\bar{n}+1)}.
    \end{equation}
    In order to compare this result to our experimental results, we need the phase-averaged version of the Husimi function, which reads
    \begin{equation}
        Q_{\mathrm{dt,inc}}=\frac{\exp{(-(|\alpha|^2 +|\alpha_0|^2)/(\bar{n}+1))}}{\pi (\bar{n}+1)} I_0\left(\frac{2 |\alpha||\alpha_0|}{\bar{n}+1}\right),
        \label{eq:IncoherentHusimi}
    \end{equation}
    where $I_0$ denotes the zeroth modified Bessel function of the first kind, $I_0(\lambda)=\frac{1}{2\pi}\int_0^{2\pi}d\phi e^{\lambda\cos{(\phi)}}$.
    This is exactly the function measured in our experiment.

    Both $Q_{\mathrm{dt}}$ and $Q_{\mathrm{dt,inc}}$ depend on exactly the same parameters, which are the coherent and incoherent photon numbers.
    Therefore, we can fit our results to $Q_{\mathrm{dt,inc}}$ to obtain these photon numbers, allowing us to determine all relevant parameters of $Q_{\mathrm{dt}}$.
    The only parameter we cannot obtain in this manner is the phase of $\alpha_0$.
    However, the quantum coherence present in the modeled displaced thermal state is exactly the same for each possible phase.

    In order to determine the photon numbers, we utilize the following relations:
    \begin{align}
        \label{eq:firstcite}\langle\hat{q}\rangle&=\sqrt{2} \mathrm{Re}(\alpha_0),\\
        \langle\hat{p}\rangle&=\sqrt{2} \mathrm{Im}(\alpha_0),\\
        \label{eq:var}\braket{(\Delta \hat{n})^2}&=|\alpha_0|^2(2\bar{n}+1)+\bar{n}^2+\bar{n},\\
        \label{eq:nc}\langle\hat{n}\rangle&=\bar{n}+|\alpha_0|^2,\\
        \label{eq:lastcite}\langle(\Delta\hat{q})^2\rangle=\langle(\Delta\hat{p})^2\rangle&=\bar{n}+\frac{1}{2}.
    \end{align}
    For pumping at the threshold, we obtain a thermal photon number below 1 and a coherent photon number that is even smaller by two orders of magnitude.
    For pumping at 7.5 $P_{\mathrm{thr}}$, we instead find a coherent photon number of $|\alpha_0|^2=53$, while the thermal photon number $\bar{n}$ amounts to about 1.7.
    The corresponding fit results are shown in the lower panels of Fig. \ref{fig:Husimis-PhaseAveraged}.
    Based on these values, we may now also determine the quantum coherence $\mathcal{C}$ of the polariton condensate for different pump powers across the threshold.
    To do so, we calculate the purities of the displaced thermal states and their incoherent counterparts as given in Eq. (\ref{eq:CohNorm}).
    We find
    \begin{equation}
        \mathrm{tr}\left(\hat{\rho}^2\right)=\frac{1}{(\bar{n}+1)^2-\bar{n}^2}
    \end{equation}
    and
    \begin{equation}
        \mathrm{tr}\left(\hat{\rho}^2_{\mathrm{inc}}\right)=\frac{\exp{(-2|\alpha_0|^2 /((\bar{n}+1)^2-\bar{n}^2))}}{(\bar{n}+1)^2-\bar{n}^2} I_0\left(\frac{2 |\alpha_0|^2}{(\bar{n}+1)^2-\bar{n}^2}\right).
    \end{equation}
    The difference yields the quantum coherence contained in the state,
    \begin{equation}
        \label{eq:Coherence}
    		\mathcal C(\hat\rho)=\frac{
    			1-
    			\exp\left[-\frac{2|\alpha_0|^2}{(\bar n+1)^2-\bar n^2}\right]
    			I_0\left[\frac{2|\alpha_0|^2}{(\bar n+1)^2-\bar n^2}\right]
    		}{(\bar n+1)^2-\bar n^2}.
    \end{equation}

\begin{figure}
    \includegraphics[width=.7\columnwidth]{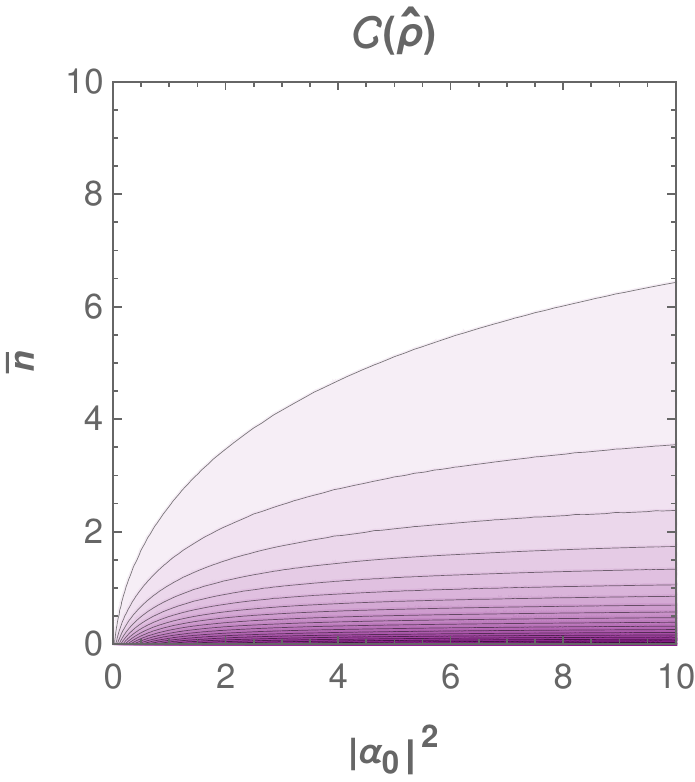}
    \includegraphics[width=.15\columnwidth]{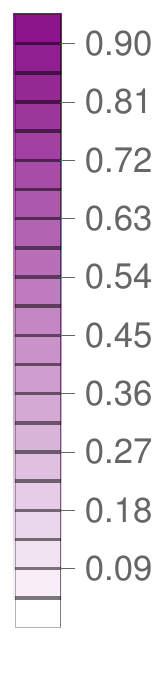}
    \caption{
        Quantum coherence $\mathcal C(\hat\rho)$ as a function of the coherent and thermal photon numbers, $0\leq|\alpha_0|^2\leq 10$ and $0\leq \bar n\leq 10$, respectively.
        With increasing displacement $\alpha_0$, the amount of quantum coherence increases.
        And a reduction of coherence is observed when the thermal background $\bar n$ increases. Reproduced from \cite{Lueders2021}.
    }\label{fig:QuantCohDTS}
\end{figure}

    Figure \ref{fig:QuantCohDTS} shows the amount of quantum coherence that is expected for different combinations of $|\alpha_0|^2$ and $\bar{n}$.
    Interestingly, even for large coherent photon numbers, already small thermal contributions are sufficient to reduce the amount of quantum coherence drastically.
    This result already implies that it is of fundamental importance to minimize residual thermal components for obtaining significant quantum coherence.

\begin{figure*}
    \includegraphics[width=0.9\textwidth]{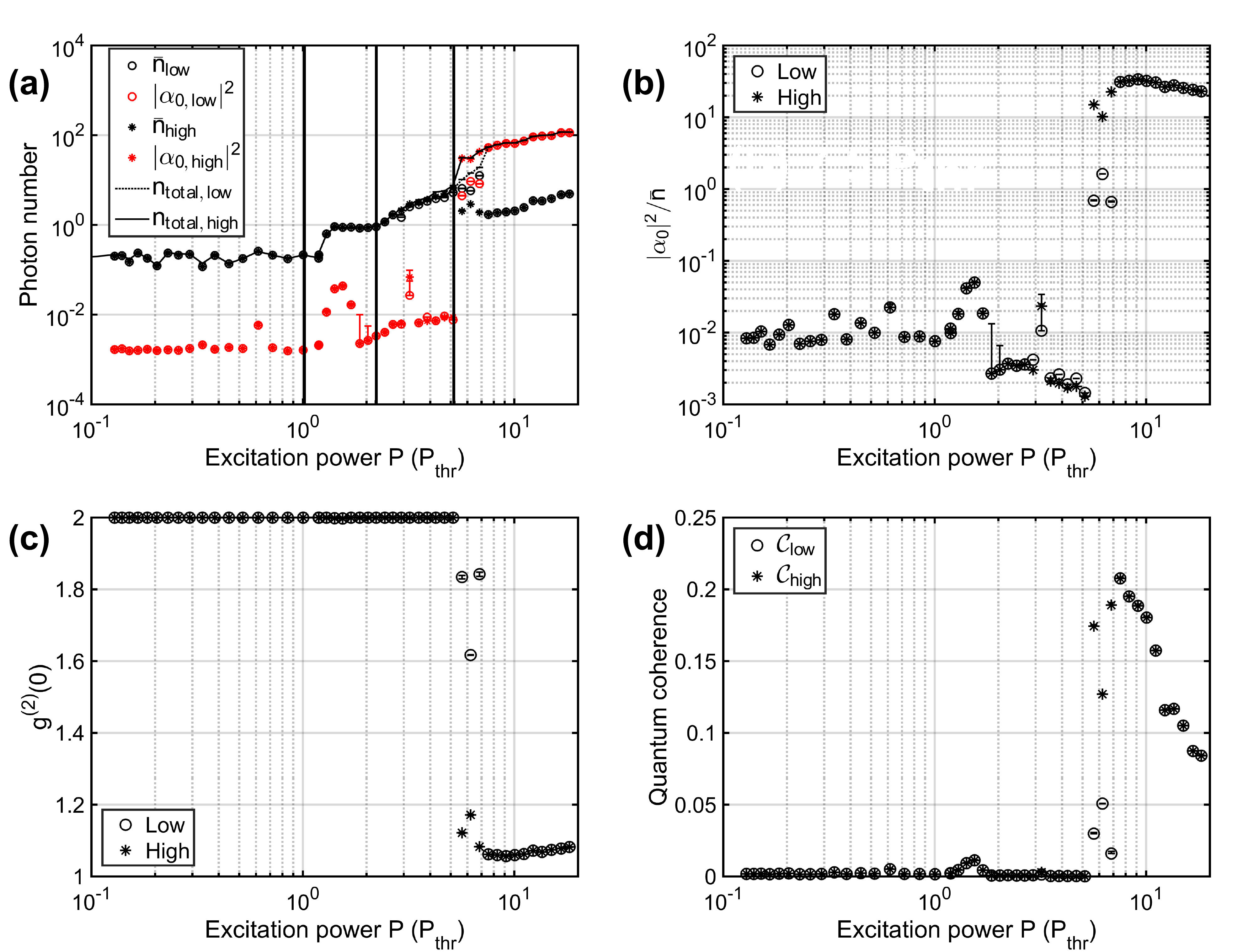}
    \caption{
        Results derived from fitting phase-averaged Husimi functions as a function of the pump intensity for vertical polarization.
        (a) Coherent (red circles) and thermal (black circles) photon numbers.
        The black line shows the total photon number $n_{\mathrm{total}}$. 
        Thick, black vertical lines indicate thresholds between different regimes of emission.
        (b) Ratio between the coherent and thermal photon number contribution.
        (c) Equal-time, second-order correlation function $g^{(2)}(\tau=0)$.
        (d) Amount of quantum coherence.
        Open and closed symbols correspond to the low and high states, respectively, when the emission is switching on and off.
        When both symbols overlap, the emission is stable.
        Error bars are obtained from a Monte Carlo error propagation;
        because of the asymmetry of the logarithmic scale, only the upper part of the error margin is depicted. Reproduced from \cite{Lueders2021}.
    }\label{fig:n-g2-Coherence}
\end{figure*}

    We performed a full pump series including pump powers from about one order of magnitude below the condensation threshold to about one order of magnitude above the condensation threshold and obtained the thermal, coherent and total photon numbers and the amount of quantum coherence present in the corresponding states.
    For comparison, we also calculated $g^{(2)}(0)$ from the same data sets; hence, we can directly compare the complementary information carried by the different coherence measures.
    The results are shown in Fig. \ref{fig:n-g2-Coherence}.
    Panel (a) shows the photon numbers.
    Panel (b) displays the ratio of the coherent to the thermal photon number.
    Panel (c) shows $g^{(2)}(0)$.
    And panel (d) indicates the amount of quantum coherence present.

    For excitation powers below threshold, the emission is almost completely thermal.
    The thermal photon number exceeds the coherent one by almost two orders of magnitude, and $g^{(2)}(0)$ takes a value of 2.
    The amount of quantum coherence present is also very close to zero.
    Taken together, the ground state polaritons are in an incoherent state and have not yet reached the required density for efficient stimulated scattering.
    Polariton relaxation is mainly governed by the thermal phonon bath and therefore incoherent \cite{Horikiri2010,Schwendimann2008}.
    Therefore, the polaritons remain in an incoherent state that does not contain superpositions.

    When the pump power reaches the threshold, both the thermal and the coherent photon numbers in the ground state increase significantly.
    In particular, the coherent photon number $|\alpha_0|^2$ rises strongly, which implies an increase of coherence.
    As the number of polaritons in the ground state approaches unity, the onset of quantum degeneracy is reached and stimulated polariton scattering towards the ground state becomes efficient.
    Notably, the quantum coherence $\mathcal{C}$ present also increases slightly but is also affected by the increase of the thermal background.

    When the pump power is increased further, the total photon numbers do not change notably up to about 2\,$P_{\mathrm{thr}}$.
    In this regime, the coherent photon number even drops and quantum coherence is lost.
    The thermal photon number exceeds the coherent one significantly, resulting in $g^{(2)}(0)=2$. 
    Within this range of powers ground state condensation competes with condensation in excited states that show up at higher momenta.
    They do not overlap with the local oscillator and are thus not recorded.
    It is, in principle, possible to study them as well by changing the properties of the local oscillator correspondingly, but the properties of the ground state are in most cases more interesting.
    We would like to emphazise that this strong mode selectivity is a significant advantage of our measurement technique.

    When increasing the pump power even further, these additional modes vanish again and the total photon number increases.
    However, the coherence still stays at a moderately low level.
    In this regime, there is significant competition between vertically and horizontally polarized condensate modes.
    We only study the vertical component here and the competition again limits the amount of coherence the condensate mode under investigation can achieve.

\begin{figure}
    \includegraphics[width=0.75\textwidth]{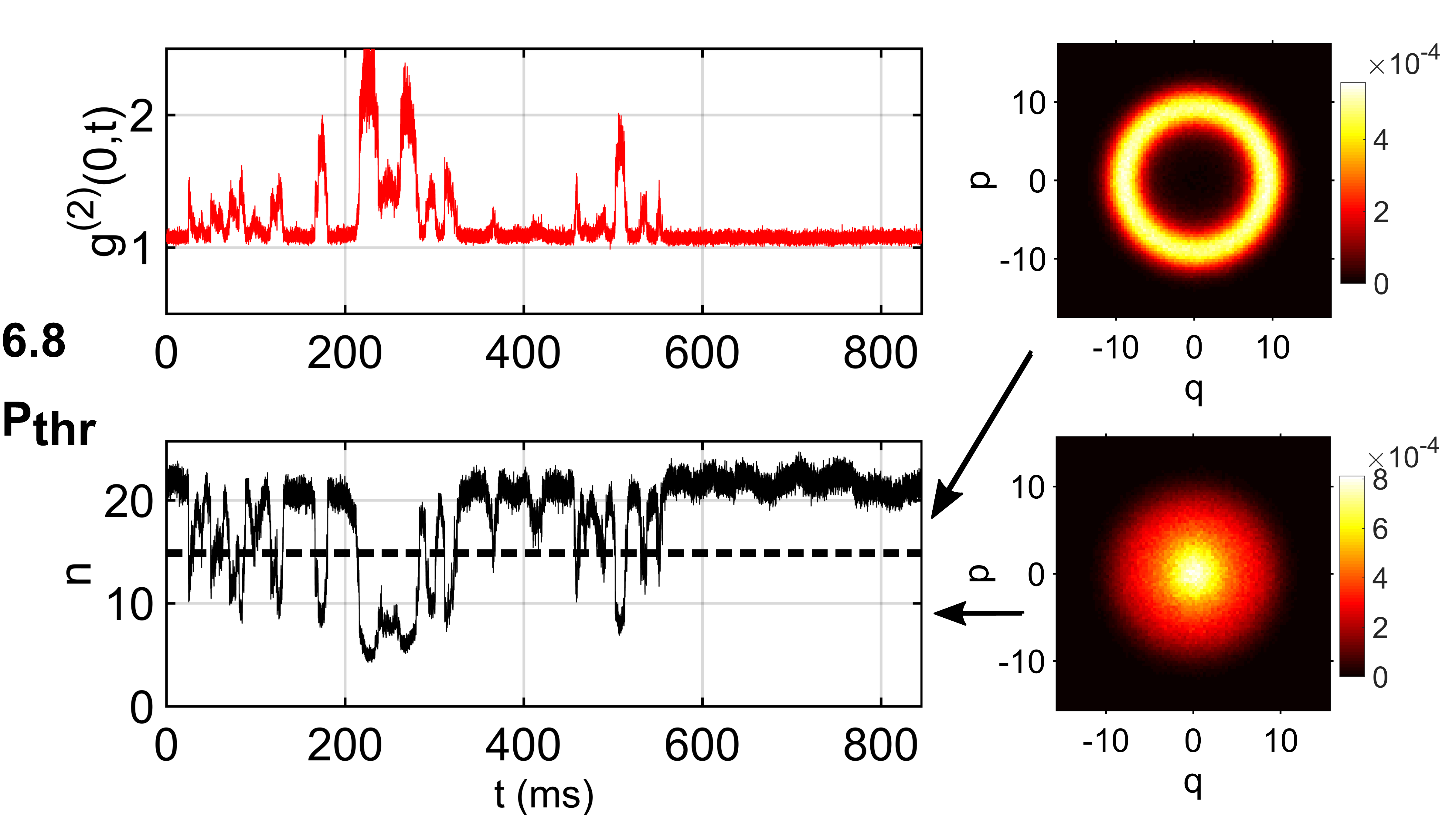}
    \caption{
        Time-resolved photon number and $g^{(2)}(\tau = 0,t)$ for $P = 6.8~P_\mathrm{thr}$.
        The initial time $t = 0$ is arbitrarily chosen as the beginning of the measurement. 
        A dashed line in the bottom panel marks the frontier between the state with high intensity and the one with low intensity.
        Right panels show the phase-averaged Husimi functions above (top plot) and below (bottom plot) this limit.
        Reproduced from \cite{Lueders2021}.
    }\label{fig:switching}
\end{figure}

    Only at much higher powers between 5\,$P_{\mathrm{thr}}$ and 7\,$P_{\mathrm{thr}}$, the coherent photon number becomes the dominant contribution.
    Here, the coherent photon number exceeds the thermal one by more than an order of magnitude, $g^{(2)}(0)$ drops towards values slightly above 1, and a significant rise in quantum coherence becomes apparent.
    In this excitation power range, ground state condensation becomes efficient.
    However, mode competition still occurs.
    When investigating the condensate emission over longer timescales on the order of one second, one can see clearly that the emission switches between states of high and low intensity as shown in Fig. \ref{fig:switching} for a pump power of 6.8\,$P_{\mathrm{thr}}$.

    The switching behavior is again a consequence of mode competition between different polarizations.
    It does not make sense physically to just average over these two scenarios.
    However, the homodyne detection spectroscopy approach allows us to perform postselective filtering.
    As every single measured quadrature is recorded individually and time-tagged, we may afterwards separate the measured data into all instances where the system is in the state of high intensity and all instances where the system is in the state of low intensity and evaluate both cases independently.
    The insets on the right of Fig. \ref{fig:switching} show that the Husimi functions for these two cases differ significantly.
    In Fig. \ref{fig:n-g2-Coherence}, we evaluate the photon numbers, $g^{(2)}(0)$ and the quantum coherence for both cases separately and provide individual values for the states of high and low intensity, respectively.

    For even higher pump powers, above 7.5\,$P_{\mathrm{thr}}$, the vertically polarized condensate mode becomes stable.
    Here, the quantum coherence of the polariton condensate reaches its maximum value of $\mathcal{C}\approx 0.21$.
    When increasing the pump power further, the coherent photon number saturates while the thermal photon number grows continuously.
    As a consequence, $g^{(2)}(0)$ increases slightly and the degree of quantum coherence drops significantly.
    This reduction of coherence at high pump powers can be explained via the increased efficiency of polariton-polariton scattering out of the ground state and detrimental sample heating which causes decoherence.
    Interestingly, the change in quantum coherence is very drastic and amounts to a reduction of more than 50$\%$, while the change in $g^{(2)}(0)$ is rather small.
    This difference again demonstrates that quantum coherence and $g^{(2)}(0)$ are complementary coherence measures carrying different information and that there is an additional benefit in characterizing the quantum coherence of a polariton condensate.

\subsection{Numerical simulations}\label{SimC}

    The dynamics of polaritons can be numerically simulated using different approaches \cite{PhysRevB.106.L220306,PhysRevResearch.3.013099}.
    Specifically, here we investigate polariton condensates that are nonresonantly excited in a planar semiconductor microcavity \cite{Kaovkinmicrocavities}.
    To describe the dynamics of such a polariton condensate $\Psi(\textbf{r})$, we use a driven-dissipative Gross-Pitaevskii equation coupled to an incoherent reservoir $n_\mathrm{res}(\textbf{r})$.
    We further apply the truncated Wigner approximation (TWA) taking into account classical and quantum fluctuations.
    Expectation values, including coherence properties, can be calculated using Monte Carlo techniques. 

    The main idea of this method is to employ the Wigner representation for a bosonic polariton field operator $\hat{\Psi}(\textbf{r})$ that can be used to sample the phase space distribution $W[\Psi(\textbf{r})]$.
    In Refs. \cite{Sperling2018, Sperling2020} comprehensive introductions are given.
    Essentially, starting with a master equation for the system's density matrix, the dynamics are transformed to the time evolution of the corresponding Wigner function via a Wigner-Weyl transformation.
    The time evolution of the Wigner function can then be truncated by neglecting third-order derivatives and mapped onto a set of stochastic partial differential equations for the corresponding complex-valued phase space variables.
    Here, $\Psi \in \mathbb{C}$ is the complex amplitude describing the polariton field \cite{PhysRevB.79.165302},
    \begin{equation}
        \label{eq:dPsi}
        d\Psi=-\frac{i}{\hbar}\Big[H+\frac{i\hbar}{2}(Rn_\mathrm{res}-\gamma_c)+g_rn_\mathrm{res}+g_c|\Psi|_-^2\Big]\Psi dt+dW\,,
    \end{equation}
    \begin{equation}
        \label{eq:dnres}
    \frac{\partial n_\mathrm{res}}{\partial t}=(-\gamma_r-R|\Psi|_-^2)n_\mathrm{res}+P_{\mathrm{pump}},  \end{equation}
    where $H=-\hbar^2\nabla^2/(2m_\mathrm{eff})$ is the free-particle Hamiltonian with the effective polariton mass $m_\mathrm{eff}=10^{-4}m_e$, with $m_e$ being the electron mass, with the decay rates $\gamma_c=0.2$~ps$^{-1}$ and $\gamma_r=1.5\gamma_c$ of the condensate and reservoir, and the interaction strength between polaritons $g_c=6\times 10^{-3}$~meV$\mu$m$^2$.
    The condensation rate and condensate-reservoir interaction are given by $R=0.015$~ps$^{-1}\mu$m$^2$ and $g_r=2g_c$.
    The values chosen for the parameters have already been used to describe the same sample in our previous work \cite{Ma2020}.
    The system is excited by a continuous-wave pump $P_{\mathrm{pump}}=P_0~\mathrm{exp}(-\textbf{r}^4/w^4)$ with width $w=65~\mu$m.
    This leads to a pump intensity for the corresponding homogeneous system of $P_{\mathrm{thr}}=\gamma_c\gamma_r/R=4$~ps$^{-1}\mu$m$^{-2}$.
    The term $|\Psi|_-^2\equiv |\Psi|^2-(\Delta V)^{-1}$ describes the renormalized condensate density, where $\Delta V=L^2/N^2$ denotes the unit-cell volumes of the two-dimensional grid. 

    Equations (\ref{eq:dPsi}) and (\ref{eq:dnres}) are solved via a fourth-order stochastic Runge-Kutta algorithm \cite{stocEq} on a finite, two-dimensional grid in real space with lengths $L=230.4~\mu$m and step size $\sqrt{\Delta V}=0.9~\mu$m, satisfying the TWA validity condition $\hbar\gamma\gg g/\Delta V$ \cite{PhysRevB.79.165302} stemming from neglecting the third-order derivative.
    For each expectation value, 300 (below threshold, 200 otherwise) realizations were evolved over a time interval of 4 ns with a fixed time step of 0.04 ps.

    The classical and quantum fluctuations within the TWA are incorporated into the condensate dynamics using complex Wiener noise $dW$, with correlations satisfying \cite{PhysRevB.79.165302}
    \begin{equation}
    \begin{split}
        \braket{dW(\textbf{r})dW(\textbf{r'})}&=0\\
        \textrm{and}~ \braket{dW(\textbf{r})dW^{\ast}(\textbf{r}')}&=(Rn_\mathrm{res}+\gamma_c)\frac{\delta_{\textbf{r},\textbf{r}'}dt}{2\Delta V}.
        \end{split}
    \end{equation}
    Expectation values of field operator products (in symmetric ordering as implied by the Wigner function \cite{VW06}) can be calculated as the average over many stochastic realizations, such as the in-cavity photon population $\hat{n}_c=\hat{b}_{\textbf{k}}^{\dagger}\hat{b}_{\textbf{k}}$ of momentum mode $\textbf{k}$ \cite{PhysRevB.76.115324},
    \begin{equation}
        \braket{\hat{n}_c}=\frac{\braket{\hat{N}}}{N_p}=\overline{|\beta_j|^2}-\frac{1}{2},
    \end{equation}
    where $\hat{b}_j\equiv \hat{b}_{\textbf{k}_j}$ with $\hat{b}_{\textbf{k}}=V^{-1/2}\Delta V\sum_{\textbf{r}}e^{-i\textbf{kr}}\hat{\Psi}(\textbf{r})$ and $\beta_j$ are the complex valued stochastic field values corresponding to operators $\hat{b}_j$.
    The corresponding variance reads
    \begin{equation}
       \braket{(\Delta \hat{n}_c)^2}=\frac{\braket{(\Delta \hat{N})^2}}{N_p}.
    \end{equation}
    The occupation is centered around $k\equiv|\textbf{k}|=0$ with finite width caused by the pump shape and the repulsive nonlinearity.
    We are taking into account this effect by averaging the polariton condensate excitation number $\hat{N}=\sum^{N_p}_{j=1}\hat{b}^{\dagger}_j\hat{b}_j$ across a pixel square of 9$\times$9 containing $N_p=81$ discrete modes centered around $k\approx 0$ \cite{PhysRevB.76.115324}.
    All expectation values are calculated after the time evolution reached the steady-state and are averaged over a period of 2~ns.
    Errors are calculated via the standard deviation and error propagation.

\begin{figure}
    \includegraphics[scale=0.435]{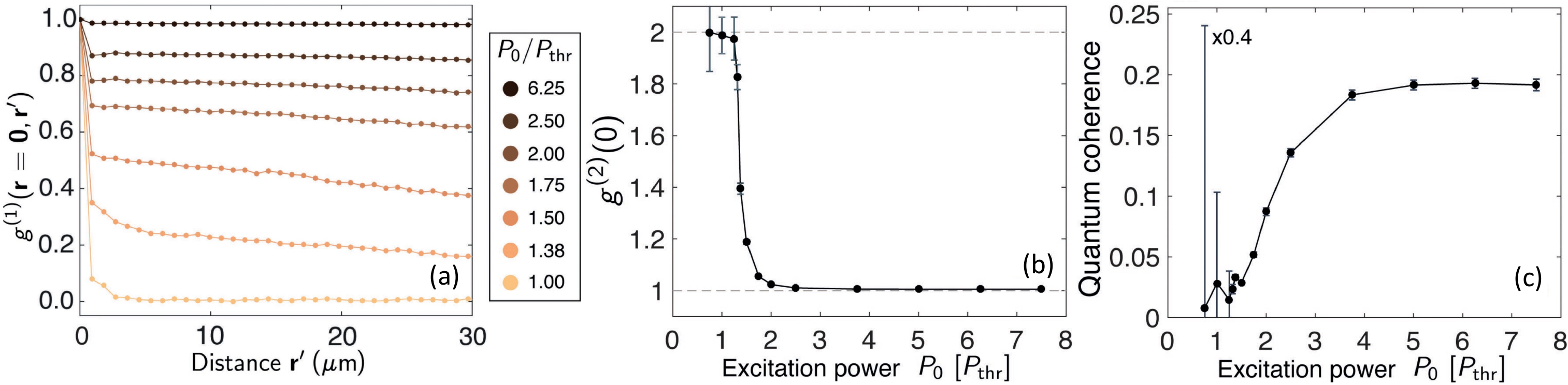}
    \caption{
        Simulated first-order, equal-time spatial coherence $g^{(1)}(\textbf{r},\textbf{r}')$ as a function of the distance from the center of the excitation spot for different pump intensities in panel (a), evidencing the creation of phase coherence across the excitation spot.
        Equal-time, second-order correlation function $g^{(2)}(0)$ showing the transition from thermal to coherent emission as a function of pump intensity in panel (b), and corresponding amount of quantum coherence $C$ in panel (c).
        Reproduced from \cite{Lueders2021}.
    }\label{fig:g1g2C}
\end{figure}

    In order to evaluate the fluctuations in the condensate occupation number, we apply the second-order, equal-time correlation function $g^{(2)}=g^{(2)}(\tau=0)$
    \begin{equation}
        g^{(2)}=\frac{\braket{\hat{n}_c^2}-\braket{\hat{n_c}}}{\braket{\hat{n}_c}^2}=2-\Bigg(1+\frac{\bar{n}}{|\alpha_0|^2}\Bigg)^{-2}
    \end{equation}
    for different pump intensities.
    Figure \ref{fig:g1g2C} clearly shows the transition from a thermal state $g^{(2)}=2$ below threshold to a coherent state with $g^{(2)}\approx1$ above threshold, justifying the treatment of the condensate state as a displaced thermal state, i.e. a thermal state which is coherently displaced in phase space, in the transitional regime.

    The polariton-number occupation for a displaced thermal state is connected to the coherent and thermal contributions $|\alpha_0|^2$ and $\bar{n}$ by the relations (\ref{eq:firstcite})-(\ref{eq:lastcite}).
    In section \ref{PatFunc} below, we explicitly reconstruct the density matrix of such states, further supporting the description of the condensate state by a displaced thermal state.

    A common measure to quantify spatial phase correlations is the first-order equal-time correlation function $g^{(1)}$, here evaluated as a function of the distance from the center of the excitation spot
    \begin{equation}
        g^{(1)}(\textbf{r},\textbf{r}')=\frac{\braket{\hat{\Psi}^{\dagger}(\textbf{r})\hat{\Psi}^{\dagger}(\textbf{r}')}}{\sqrt{\braket{\hat{\Psi}^{\dagger}(\textbf{r})\hat{\Psi}(\textbf{r})}\braket{\hat{\Psi}^{\dagger}(\textbf{r}')\hat{\Psi}(\textbf{r}')}}}.
    \end{equation}
    In Fig. \ref{fig:g1g2C} (a) this first-order correlation is depicted for different pump intensities showing a buildup of spatial phase coherence in the polariton condensate for increasing intensities, becoming almost uniform with $g^{(1)}\approx1$ across the excitation spot for $P_0=6.25~P_\mathrm{thr}$.

    However, $g^{(1)}$ is a macroscopic quantitative measure for the length scale over which phase correlations are preserved and cannot provide additional information about the quantum nature of a state.
    Moreover, even though $g^{(2)}$ gives information about the excitation number distribution of a state, it cannot specify its course nor the off-diagonal contributions of the density matrix.

    Here, we determine the measure of quantum coherence evaluated via Eq. (\ref{eq:Coherence}) as the amount of quantum superpositions of particles, as described in section \ref{QCDefinition}, by assessing the off-diagonal contributions in the density operator in Fock space.
    In Fig. \ref{fig:g1g2C}, the quantum coherence is shown as a function of the pump intensity.
    In fact, we are able to quantify a nonvanishing amount of quantum coherence as a result of quantum superposition of polaritons, evident as a significant increase of quantum coherence across the threshold that saturates in a plateau at around $\mathcal{C}\approx 0.2$ which is in good agreement with the experiment.

    In Fig. \ref{fig:g1g2C} (b), the $g^{(2)}$ function shows the transition to $g^{(2)}\approx 1$, i.e. a coherent state, already at around twice the threshold power while the quantum coherence saturates for higher values.
    As mentioned in the previous Sec. \ref{SteadyQC}, the quantum coherence and $g^{(2)}$ are complementary coherence measures.
    In general, the $g^{(2)}$ function does not provide information on the quantum nature of matter systems and cannot quantify the amount of quantum coherence, focusing on very different aspects that pertain to classical waves and non-classical quantized light where particles are deemed classical.
    Moreover, it cannot unambiguously classify a state.
    The quantum coherence, however, directly quantifies the amount of quantum superpositions of particles providing additional information on the system.

    While the actual value of $\mathcal{C}\approx 0.21$ found for the polariton condensate is lower than the ideal value of 1, it still indicates a significant amount of quantum superpositions present in the condensate, which may be used for quantum technological tasks.
    For example, direct conversion of quantum coherence into entanglement or other types of non-classical correlations is possible \cite{Killoran2016,Chitambar2016,Qiao2018,Ma2016}.
    Accordingly, quantum coherence is a quantum computational resource that is operationally equivalent to entanglement \cite{Streltsov2017,Chitambar2019,Winter2016} and therefore highly interesting for implementing protocols in quantum information.
    In practice, this means that it is possible to transform the superposition of Fock states into entanglement using incoherent processes, where the amount of entanglement created is identical to the amount of quantum coherence present before the conversion \cite{Killoran2016}.
    Parametric pair production processes are typical processes that may be utilized to implement this conversion.
    In polariton systems, similar parametric processes exist, e.g., for the polariton optical parametric oscillator \cite{Baumberg2000,Amo2009,Spano2012} or in quantum depletion \cite{Pieczarka2020}. 

    In summary, we applied the concept of quantum coherence to demonstrate the general resourcefulness of the generated polariton states in quantum protocols for quantum processing tasks within the framework of quantum resource theories.
    We did not implement any explicit protocol, which will be an interesting task for the future.
    However, the quantitative measure of quantum coherence we determine allows to estimate the performance of polariton systems in such tasks and also represents a suitable performance identifier for optimizing polariton systems. 

\section{Conditional measurements and postselective spectroscopy}

    In the previous section, we established quantum coherence as a valuable resource for quantum technologies and evaluated the amount of quantum coherence contained in a polariton condensate at a given instant.
    However, it is even more important to determine for how long this coherence may be preserved.
    In the following, we show that time-resolved postselective conditional homodyne spectroscopy enables us to do exactly that.

\begin{figure}
    \includegraphics[width=0.65\textwidth]{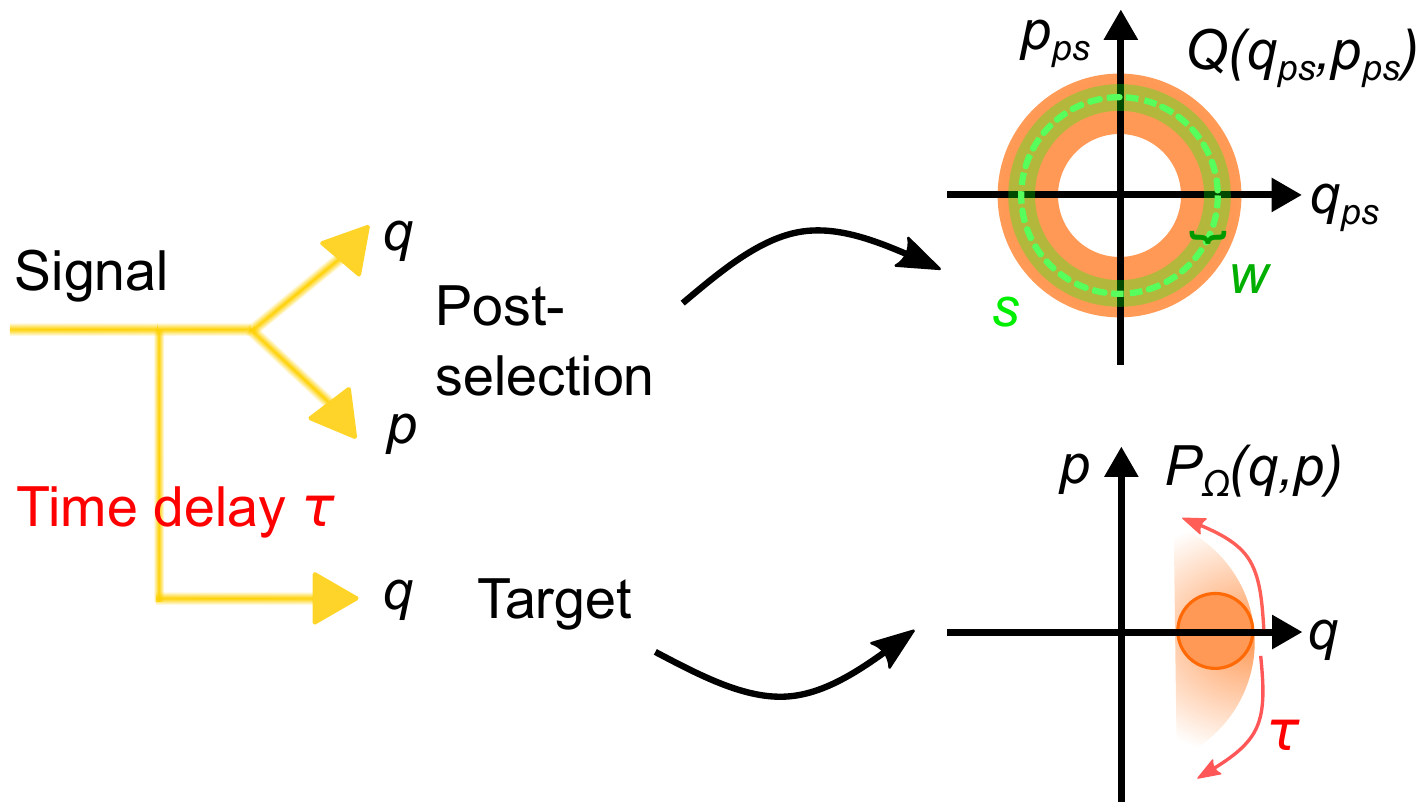}
    \caption{ 
        Experimental concept of postselective conditional spectroscopy.
        We split the signal into three homodyne detection channels.
        Two are used for intensity selection, delivering field quadratures $(q_\mathrm{ps}, p_\mathrm{ps})$ of a Husimi function $Q(\alpha_\mathrm{ps})$, with $\alpha_\mathrm{ps}=q_\mathrm{ps}+ip_\mathrm{ps}$.
        A subset of detection events corresponding to a specific region with radius $s$ and width $w$ in phase space is chosen, fixing the postselected signal's mean intensity and its uncertainty.
        The third quadrature is the target channel for quadratures $q$ at some delay $\tau$ that determines the elapsed time of the system's evolution.
        Together with the corresponding phases, this allows us to directly sample $P_\Omega$.
        Figure reproduced from \cite{Lueders2023}.
    }\label{Fig:idea}
\end{figure}

    Using the same sample and excitation conditions as in the previous sections, we perform time-resolved tomography on the emission by dividing the signal into three homodyne detection channels as shown in Fig. \ref{Fig:idea}.
    Two of these channels are the postselection arm which provides a proxy measurement of the Husimi Q function \cite{Lueders2021}, while the third one constitutes the target arm \cite{Thewes2020a}.
    The temporal offset $\tau$ between the first two channels and the third one is controlled via a delay line.
    Any single measurement result in the first two channels consisting of the quadratures $q_\text{ps}$ and $p_\text{ps}$ provides an estimate for the instantaneous amplitude and phase of the signal light field.
    As all of the LOs used to have a well-defined phase relation with respect to each other, we may reconstruct the relative phase $\varphi$ between the signal and the LO in the target channel by adding the relative phase $\Delta \varphi$ between the local oscillators in the postselection arm and the target arm to the signal phase $\varphi_\text{ps}=\arctan(p_\text{ps}/q_\text{ps})$ measured in the postselection arm.
    All possible values of $\varphi$ are sampled if the measurement duration is long enough.
    We obtain sets $(\varphi,q)$ of phases and corresponding quadratures in the target arm, which we utilize to construct a phase space function that provides information about the dephasing of the signal.

\subsection{Regularized phase-space functions}

    In order to track the coherence produced by the polariton condensate in time, we adopt advanced phase space functions denoted as $P_\Omega$.
    While the Glauber-Sudarshan distribution $P$ as defined in Eq. (\ref{eq:GS}) is the prototypical phase-space distribution based on coherent fields, it may be ill-behaved mathematically and exhibit an exponential order of singularities \cite{Sperling2016}.
    In order to avoid these problems, convolution-based regularization procedures were developed \cite{Cahill1969,Agarwal1970}, which read
    \begin{equation}
        \label{eq:Pconvo}
        P_\Omega(\alpha)=\int d^2 \gamma \Omega(\gamma-\alpha)P(\gamma).
    \end{equation}
    When using a Gaussian kernel for $\Omega$, one may for example obtain the Wigner function and the Husimi $Q$ function.
    Such regularized phase space functions have been utilized in pioneering works in semiconductor spectroscopy \cite{Kira2008,Kira2011,Almand-Hunter2014}.
    However, the reconstruction of these common phase space functions may still be challenging due to ill-posed inversions \cite{Tan1997,Starkov2009}, diverging pattern functions \cite{LEONHARDT1996144,RICHTER1996327} or highly non-trivial maximum likelihood estimations \cite{Hradil1997,Lvovsky2004}.
    Instead, it has been proposed and demonstrated that tailored non-Gaussian kernels may be highly beneficial for characterizing light fields in phase space \cite{Kiesel2010,Kuehn2021,Koehnke2021}.

    In order to construct a suitable phase space function for the experimental data we obtained, we use the following non-Gaussian, phase-invariant kernel:
    \begin{equation}
        \Omega(\gamma)=\left(\frac{J_1(2 R|\gamma|)}{\sqrt{\pi}|\gamma|}\right)^2,
        \label{eq:Omega_kernel}
    \end{equation}
    where $J_1$ corresponds to the first Bessel function of the first kind and $R$ is an adjustable width parameter.
    In our case, we find good results for a value of $R=0.7$.
    With this approach, $P_\Omega$ can be directly obtained from the experimental data \cite{Lueders2023}, avoiding the problems outlined above \cite{Kiesel2011}.

    All incoherent contributions contained in the diagonal elements of the density operator depend on the amplitude $|\alpha|$ of $P_\Omega$ only, but not on its phase $\phi$.
    On the other hand, the coherent contributions contained in the off-diagonal elements of the density operator directly relate to $\phi$, but not to $|\alpha|$.
    Accordingly, the angular width of $P_\Omega$ in phase space is a reasonable operational measure of quantum coherence.
    In Sec. \ref{QCDefinition}, we outlined the idea of this approach for the non-regular $P$ distribution.
    Here, we choose the circular variance \cite{Fisher1993}
    \begin{equation}
        \text{Var}(\phi)=1-|\langle r \rangle|, \quad \text{with} \quad\langle r \rangle=\int d^2 \alpha P_\Omega(\alpha)\frac{\alpha}{|\alpha|} 
        \label{eq:PhaseVariance}
    \end{equation}
    for quantifying quantum coherence in terms of the regualrized $P_\Omega$, which takes its maximal value of 1 for a fully incoherent and therefore phase-randomized state.
    A high degree of quantum coherence is obtained for a narrow phase distribution.
    Thus, a low circular variance indicates the presence of substantial amounts of quantum coherence.
    To monitor its decay dynamics, in the following, we track the temporal evolution of the circular variance of $P_\Omega$.

    To this end, we apply postselective conditional spectroscopy.
    We consider the experimental data, which consists of the quadratures $q_\text{ps}$ and $p_\text{ps}$ measured in the postselection arm and the corresponding pairs of quadratures and relative phases $(\varphi,q)$ measured in the target arm at a time delay $\tau$.
    The former data provide an estimate of the instantaneous amplitude and phase of the light field, while the latter data allow us to gain information about the light field at a delay $\tau$.

\begin{figure*}
    \includegraphics[width=1\textwidth]{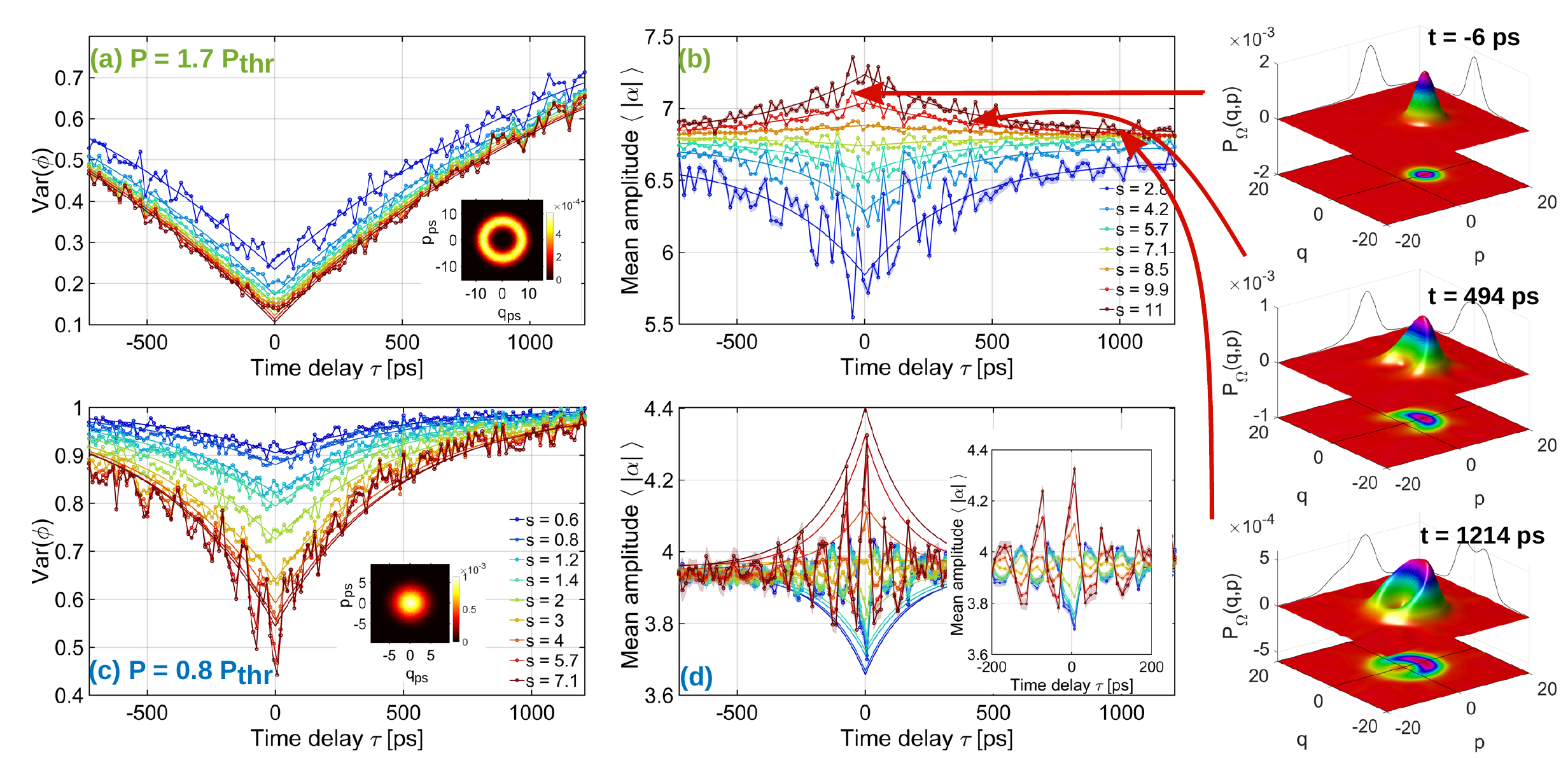}
    \caption{
        Temporal behavior of the mean amplitude $\langle |\alpha|\rangle$ and the circular phase variance $\mathrm{Var}(\phi)$ of reconstructed $P_\Omega$, depending on the selected radius $s$.
        A shaded area around the curves corresponds to a one standard deviation error margin, directly derived from $P_\Omega$'s uncertainty, but is mostly not visible.  
        Lines depict exponential fits.
        Plots (a) and (b) are for $P_{\text{exc}} = 1.7~P_\mathrm{thr}$.
        The width $w$ of the selected phase space region is $0.57$.
        Plots (c) and (d) are for $P_{\text{exc}} = 0.8~P_\mathrm{thr}$.
        The width $w$ is $0.1$, except for the two highest radii $s$, where $w = 0.57$. 
        Insets in (a) and (c) show Husimi functions from which the radius $s$ is selected, cf. Fig. \ref{Fig:idea}.
        The inset in (d) displays a zoom of the region $\tau\approx 0$, revealing an oscillation of the mean amplitude.
        Right column exemplifies three $P_\Omega$ for $P_{\text{exc}} = 1.7~P_\mathrm{thr}$ and a selected radius $s = 9.9$ at time delays $\tau = -6~\mathrm{ps}$ (top), $\tau = 494~\mathrm{ps}$ (middle), and $\tau = 1\,214~\mathrm{ps}$ (bottom).
        Figure reproduced from \cite{Lueders2023}.
    }\label{Fig:delayplots}
\end{figure*}

    In order to obtain the desired dynamics of quantum coherence, we apply postselection to the former data.
    As shown in Fig. \ref{Fig:idea}, we construct the Husimi $Q$ function of the light field in the postselection arm from $q_\text{ps}$ and $p_\text{ps}$ and postselect on all detection events that fall within an annulus-shaped region specified by a radius $s$ and a thickness $w$ in phase space. 
    This is equivalent to choosing a certain range of light field amplitudes.
    We now identify all pairs $(\varphi,q)$ in the target arm that correspond to the selected range in the postselection arm.
    We want to emphasize that $\varphi$ is a relative phase defined with respect to the phase measured in the postselection arm.
    We then reconstruct $P_\Omega$ from these postselected datasets for different delays $\tau$. 
    Figuratively speaking, this answers the question:
    Given we found some certain amplitude and phase for the light field in the postselection arm at some well-defined time, what does the distribution of amplitudes and phases in the target arm look like at some delay?

    For short delays, we expect both to show a narrow distribution governed by the postselection conditions while, on longer timescales, both will change.
    If the postselected amplitude does not correspond to the mean modulus of the amplitude, one witnesses relaxation towards the mean amplitude.
    Because of dephasing, also the relative phase begins to undergo diffusion in phase space, resulting in a larger circular variance.
    The former process involves the incoherent contributions to the density operator, while the latter involves the coherent contributions.
    The huge advantage of continuous variable spectroscopy in phase space lies in the possibility to unravel these two contributions and to study them separately.

    The right column of Fig. \ref{Fig:delayplots} shows reconstructed $P_\Omega$ functions for three characteristic delays at an excitation power of 1.7 times the threshold excitation power.
    The phase diffuses with increasing $\tau$ and simultaneously the mean amplitude relaxes towards the steady-state value.
    In order to quantitatively assess the dynamics, we calculate the mean amplitude $\langle|\alpha|\rangle=\int d^2 \alpha P_\Omega(\alpha)|\alpha|$ and the circular phase variance as given by Eq. (\ref{eq:PhaseVariance}) for different delays $\tau$ and postselected amplitudes $s$.
    The results for a constant width $w=0.6$ of the postselection radius are shown in panels (a) and (b) of Fig. \ref{Fig:delayplots}.
    Solid lines in the figure correspond to exponential fits to the data.

    We find that $\text{Var}(\varphi)$ takes its minimum around $\tau=0$ with a minimal value of 0.14 for the highest postselection amplitude $s=11$.
    This minimal value increases when postselecting on smaller values of $s$.
    Smaller values of $s$ effectively correspond to a lower number of polaritons being present, so the reduction in quantum coherence may be readily explained by the phase-photon number uncertainty relation.
    For longer delays, the circular variance increases, but a fully uniform distribution corresponding to $\text{Var}(\varphi)=1$ is not reached within the range of delays investigated here, which goes up to 1.2\,ns.
    In contrast, compared to the quantum phase, the mean amplitude $\langle|\alpha|\rangle$ decays much faster and relaxes almost completely towards the steady-state value.

    Panels (c) and (d) of Fig. \ref{Fig:delayplots} show the results for a lower excitation power close to the threshold.
    Here, $\text{Var}(\varphi)$ increases significantly faster and reaches almost full phase decoherence.
    The mean amplitude $\langle|\alpha|\rangle$ rapidly relaxes towards the steady state. 
    Interestingly, we are also able to identify amplitude oscillations at a frequency of about 12.5\,GHz in this excitation power range.
    We would like to point out that we are able to resolve these spontaneous oscillations due to the postselection procedure we apply.
    These dynamics are not observable when only the steady state of the system is studied.
    Within this frequency range, there are two possible causes of this effect.
    First, the condensate may support two modes of orthogonal polarizations.
    For not too strong pumping, mode competition between these modes may occur and result in effectively bistable or oscillatory behavior \cite{Sigurdsson2020}.
    Alternatively, also spatial density modulations of the polariton condensate, such as breathing modes may appear in this frequency range \cite{Estrecho2021}.

\begin{figure*}
    \includegraphics[width=0.8\textwidth]{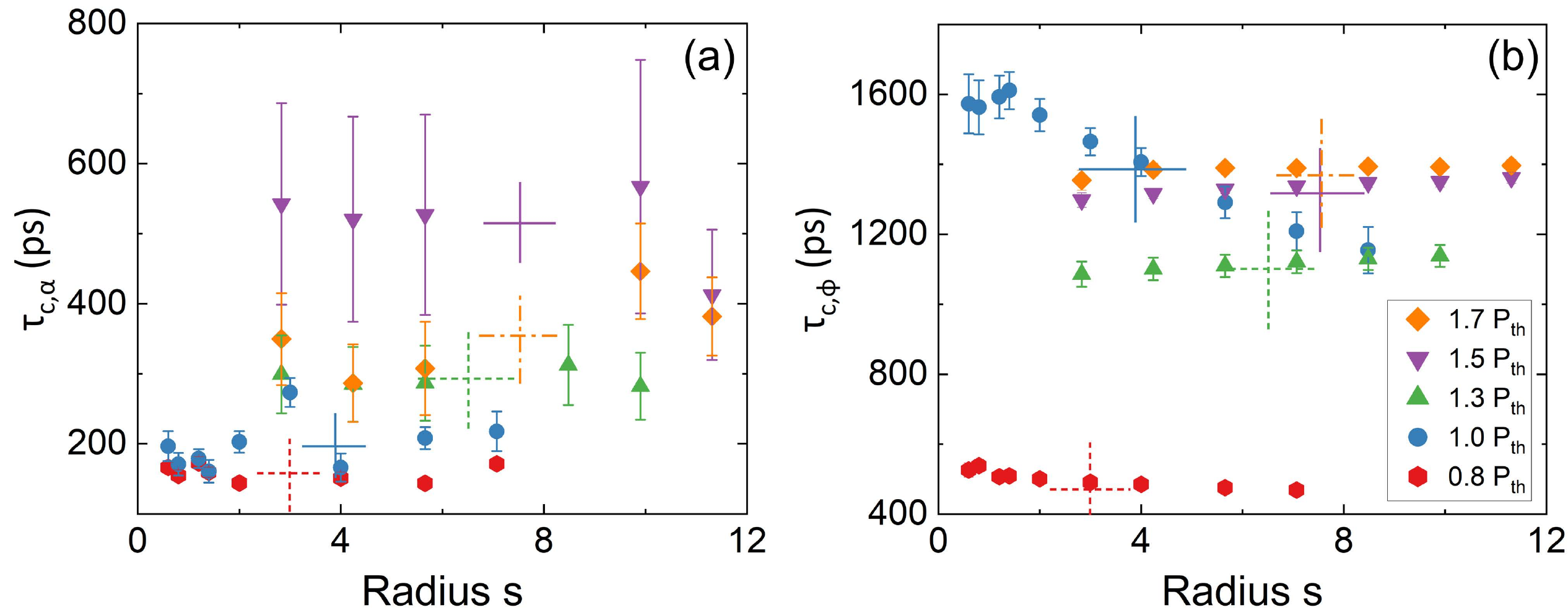}
    \caption{
        (a) Amplitude and (b) phase decay times as a function of the selected radius s for different excitation powers.
        Crosses indicate the mean amplitudes (vertical line) and decay times (horizontal line) of the stationary state, where the mean radius is averaged over the Husimi function.
    }\label{fig:decays}
\end{figure*}

    We may also provide quantitative estimates for the amplitude and phase decay times $\tau_{c,\alpha}$ and $\tau_{c,\varphi}$, respectively.
    For this purpose, we perform exponential fits
    \begin{align}
        \langle|\alpha|\rangle(\tau)&=\langle|\alpha|\rangle_\text{steady} + \Delta \langle|\alpha|\rangle \exp(-\tau/\tau_{c,\alpha})\\
        \text{Var}(\varphi,\tau)&=1-\Delta \text{Var}(\varphi) \exp(-\tau/\tau_{c,\varphi}).\label{eq:fitcv}
    \end{align}
    The obtained decay times are shown in Fig. \ref{fig:decays}.
    For the amplitude, we find no strong dependence on $s$.
    However, we find some dependence on the pump power.
    $\tau_{c,\alpha}$ increases from values below 200\,ps to values above 500\,ps for pumping at 1.5 times the threshold pump power and then decreases again.
    It is important to note that long decay times of the conditional amplitude do not necessarily correspond to enhanced coherence.
    Due to the nature of the conditional measurement, we get little information about our signal when the postselected amplitude is close to the mean amplitude in the target arm. In that case, the amplitude in the target arm does not change much with time.
    Only if the postselected amplitude deviates from the steady state amplitude, we observe resolvable amplitude dynamics.
    Effectively, these dynamics correspond to how quickly a perturbation in amplitude becomes damped out.
    Long damping times then obviously do not necessarily correspond to enhanced coherence. 
    On the contrary, longer decay times of $\text{Var}(\varphi)$ indeed indicate longer persistence of quantum coherence.
    Here, for pumping above threshold we find a clear tendency for states postselected on higher $s$ to have a longer decay time, while we find an opposite trend for pumping close to the threshold.
    This might be explained by considering that at low pump powers mode competition is a relevant effect and the system is less stable against perturbations.
    For pumping exactly at the threshold, the phase decay time becomes exceptionally long.
    In this range, the system is only marginally affected by heating, interaction with the reservoir and nonlinear polariton-polariton interactions which become relevant sources of decoherence at higher pump powers.

    We emphasize that the postselected phase coherence times we observe are surprisingly long.
    They range somewhere between 520\,ps and 1390\,ps for different pump powers.
    Our result is in line with recent observations of long coherence times on the order of 1\,ns for experiments, where the spatial overlap between the polariton condensate and the reservoir is minimized \cite{Orfanakis2021,Baryshev2022}.
    However, in our case, there is a significant spatial overlap between condensate and reservoir. 
    Under such conditions, coherence times are usually limited to about 100\,ps, even when using low-noise single-mode lasers and spatially confined single-mode cavities \,ps \cite{Love2008,Kim2016}. 
    Interestingly, most of these studies relied on classical first-order temporal coherence measurements to determine the coherence time of the condensate and found a Gaussian shape for $g^{(1)}(\tau)$, especially at higher pump powers \cite{Kim2016}.
    This shape implies the presence of inhomogeneous spectral broadening.

    We further emphasize that our experimental approach differs from such measurements in several respects.
    First, the local oscillator used in homodyne detection acts as a spectral mode filter and therefore effectively removes contributions from other modes which might result in inhomogeneous broadening for incoherent detection.
    Second, due to the postselection process, the initial reference state of the light field at time 0 is always well-defined and not averaged over many possible initial states as is the case in a measurement of $g^{(1)}(\tau)$.
    Third, we are able to consider phase and amplitude relaxation separately, while a measurement of $g^{(1)}(\tau)$ necessarily includes both.
    Accordingly, we are able to minimize the influence of inhomogeneous broadening, which is of fundamental interest for spectroscopic studies in materials science.

\subsection{Theory of decay of quantum coherence}

    In Section \ref{SteadyQC}, the spatial phase coherence and quantum coherence of a polariton condensate were studied.
    A common method to study phase coherence is the first-order unequal-time correlation function $g^{(1)}(\tau)=\braket{\hat{a}^\dagger(t)\hat{a}(t+\tau)}$.
    Within the here applied truncated Wigner framework however, $g^{(1)}(\tau)$ cannot be directly calculated because unknown unequal-time commutators are needed for the evaluation of $\braket{\hat{a}^\dagger(t)\hat{a}(t+\tau)}$ \cite{PhysRevA.80.033624}.
    Instead, we apply the circular variance of $P_{\Omega}$ defined by Eq. (\ref{eq:PhaseVariance}) \cite{Lueders2023,pukropDiss}.   

    To study the temporal decay of quantum coherence in the numerical simulation, the initial condition of the time evolution has to be prepared.
    Otherwise, if an empty cavity is assumed as initial condition, all numerical samples have randomized phases, resulting in $\braket{\alpha(t)}=0$ at all times.
    This leads to vanishing off-diagonal density matrix elements, i.e. $\mathcal C=0$.
    Moreover, the specific result Eq. (\ref{eq:Coherence}) is only applicable to states with Gaussian phase distribution.

    Therefore, we start with a displaced thermal state with Gaussian distribution in phase space as initial condition.
    This means that the mean and circular variance of the phase angle are fixed and then evolved in time.
    Specifically, the displaced thermal state is prepared in a way that it has the same mean occupation number and quantum coherence as the steady-state.
    This way, the fluctuations in amplitude and phase are only determined by the dynamical noise and are not influenced by randomization of the initial states.

    The aim of the initial state preparation is to create a phase coherent state with Gaussian phase space distribution that meets the values for the expectation values $\braket{\hat{n}},\braket{(\Delta \hat{n})^2}$ and resulting quantum coherence $C$ of the desired state \cite{pukropDiss}.
    These values are determined through the corresponding TWA steady-state calculation created with vacuum initial condition; see Sec. \ref{SimC}.
    The first step is to normalize the stationary mean-field model condensate solution, i.e. $dW=0$ and $|\Psi|^2_-=|\Psi|^2$, creating the spatial envelope of the initial state.
    Then, $N_s=2000$ phase space samples are drawn from a Gaussian distribution with mean $\mu$ and standard deviation $\sigma$ for each spatial grid point.
    The normalized spatial envelope including the phase is then multiplied with the grid.
    The $N_s$ samples are used to determine the expectation values $\braket{\hat{n}},\braket{(\Delta \hat{n})^2}$ and resulting quantum coherence $C$ which are then compared to the values of the corresponding TWA steady-state calculation.
    This process is repeated for different values for $\mu$ and $\sigma$ until a sufficient agreement with the three target values is reached.
    We note that this method yields perfect results for the phase space and expectation values, however, the steady-state maximum value of the real-space density is typically slightly overestimated, leading to initial density oscillations while the system dynamically approaches the steady-state target density.
    With an initial time evolution for 25~ps with reduced decoherence effects, i.e. $g_c=0$, phase coherence is preserved while the maximum real space density comes close to the steady-state solution. 

    This initial condition with desired phase coherence (instead of random phases) and correct density is then evolved in time according to Eqs. (\ref{eq:dPsi}) and (\ref{eq:dnres}) and decoheres over time.
    The initial reservoir state is taken from the stationary mean-field solution.
    The excitation pump is changed to fit to the FWHM of the experimental pump with $P_{\mathrm{pump}}=P_0~\mathrm{exp}(-\textbf{r}^2/w^2)$, width $w=40~\mu$m, and $P_{\mathrm{thr}}=8$~ps$^{-1}\mu$m$^{-2}$, this time defined analogously to the experiment by the onset of quantum coherence in the selected mode \cite{Lueders2023}.

    By evolving the initial state according to Eqs. (\ref{eq:dPsi}) and (\ref{eq:dnres}) within the TWA approach we sample the Wigner function of the system, whereas in the experiment the Husimi function is sampled.
    Still, by employing a convolution-deconvolution approach giving a relation between the Wigner function and the here applied $P_\Omega$, we are able to define an approach compatible with experiment.
    Analogously to Eq. (\ref{eq:Pconvo}), where the Husimi function, i.e. the experimental data, is convoluted with the non-Gaussian, phase-invariant kernel in Eq. (\ref{eq:Omega_kernel}), $P_{\Omega}(\alpha)$ can be obtained by carrying out the convolution of the Wigner function with the kernel $K(\alpha)$.
    In Fourier space, the transformation from the Wigner function $\widetilde W(\beta)$ to the Glauber-Sudarshan distribution $\widetilde P(\beta)$ is given by $\widetilde P(\beta)=e^{|\beta|^2/2}\widetilde W(\beta)$.
    The transformation from $\widetilde P(\beta)$ to the characteristic function $\widetilde P_\Omega$ of the regularized phase space function can be described via the autocorrelation function of a (normalized) cylinder, i.e., the Fourier transform of the kernel in Eq. \eqref{eq:Omega_kernel}. 
    For comparison with  experiment we also choose $R=0.7$ in the numerical simulation.
    Combining these relations and computing the inverse Fourier transform yields the convolution kernel $K(\alpha)$ that maps the Wigner function $W(\alpha)$ to $P_\Omega$ \cite{Lueders2023},
	\begin{equation}
	\begin{aligned}
		K(\alpha)
		=&\int \frac{d^2\beta}{\pi^2}e^{\beta^\ast\alpha-\beta\alpha^\ast}e^{|\beta|^2/2}
		\int \frac{d^2\gamma}{\pi R^2}\Theta(R-|\gamma|)\theta(R-|\gamma+\beta|)
		\\
		=&\frac{1}{\pi^3 R^2}\int_{|\gamma|\leq R} d^2\gamma\int_{|\delta|\leq R}d^2\delta\,
		\exp\left[
			(\delta-\gamma)^\ast\alpha
			-(\delta-\gamma)\alpha^\ast
			+\frac{|\delta-\gamma|^2}{2}
		\right]
		\\
		=&\frac{1}{\pi^3 R^2}\int_0^R ds\,s\int_0^R dt\,t \int_0^{2\pi}d\varphi\int_0^{2\pi}d\vartheta\,
		e^D
		\\
		=&\frac{R^2}{\pi^3}\int_0^1 ds's'\int_0^1 dt't' \int_0^{2\pi}d\varphi\int_0^{2\pi}d\vartheta\,
		e^E,
	\end{aligned}
	\end{equation}
	where
	\begin{equation}
	\begin{aligned}
			D=-2i|\alpha|t\sin(\vartheta)+2i|\alpha|s\sin(\varphi)+\frac{1}{2}\left(s^2+t^2-2st\cos(\varphi-\vartheta)\right)~\textrm{and}\\
			E=-iA\Big(t'\sin(\vartheta)-s'\sin(\varphi)\Big)+\frac{R^2}{2}\Big(t^{\prime 2}+s^{\prime 2}-2s't'\cos(\varphi-\vartheta)\Big),
	\end{aligned}
	\end{equation}
    using the substitutions $\delta=\gamma+\beta$, $\gamma=s e^{i(\varphi+\arg\alpha)}$, $\delta=t e^{i(\vartheta+\arg\alpha)}$, $s'=s/R$, $t'=t/R$, and $A=2R|\alpha|$ during the derivation.
    The resulting integral for $K(\alpha)$ over finite domains can be evaluated numerically, e.g. by using Monte Carlo integration over all TWA realizations.
    The circular variance can then be calculated according to Eq. (\ref{eq:PhaseVariance}) and the coherence time $\tau_c$ evaluated via the exponential fit in Eq. (\ref{eq:fitcv}).

\begin{figure*}
    \includegraphics[scale=0.43]{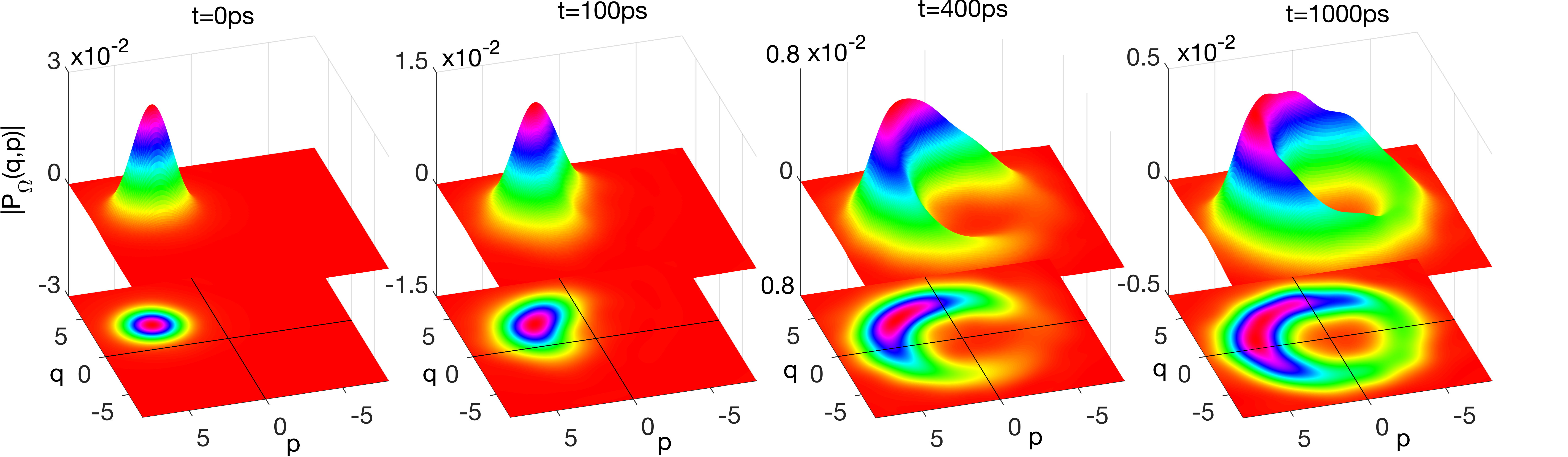}
    \caption{
        Numerical results for the time evolution of the regularized P-function $P_{\Omega}$ of the state for $P=1.25P_{\textrm{th}}$.
    }\label{fig:numP}
\end{figure*}

    In the experiment, the coherence time saturates for high pump powers due to decoherence mechanisms like temperature effects (heating of the sample) or higher-order scattering processes.
    We cannot observe this trend in the numerical simulation since those mechanisms are not included in the model.
    Hence, the coherence times in the simulation show an ongoing increase for higher pump powers, with numerically determined coherence times in good agreement with the experimental values around the threshold pump power \cite{Lueders2023}.
    The decoherence in the numerical model is caused by the effective potential whose nonlinear part up to first order reads $V_{\mathrm{nl}}= g_c[1-P_{\mathrm{pump}}^*(g_r\gamma_c)/(g_c\gamma_r)]|\psi|^2$ \cite{PhysRevLett.113.200404}, with $P_{\mathrm{pump}}^{*}\equiv P_{\mathrm{pump}}R_r /(\gamma_c\gamma_r)$.
    Trends for $\tau_c$ are also analyzed by our numerical calculations, showing that in accordance with the effective potential $\tau_c$ decreases with increasing interaction strength $g_c$ and with decreasing condensate-reservoir interaction $g_r$.

    The regularized $P$ functions $P_{\Omega}(\alpha)$ evaluated for the zero-momentum mode for different points in time for the condensate state at $P=1.25P_{\textrm{th}}$ are depicted in Fig. \ref{fig:numP}.
    The first panel shows the regularized $P$ function of the initial state which has been prepared and optimized until it is in sufficient agreement with the values for the quantum coherence $\mathcal C=0.24$, expectation value $\braket{\hat{n}}=24.4$, and variance $\braket{(\Delta\hat{n})^2}=89.9$ of the steady-state solution.
    The resulting $P_{\Omega}$ representation has a localized Gaussian shape.
    As time progresses, the condensate state relaxes towards a decoherent state and becomes less Gaussian and more annulus shaped while its amplitude decreases.

\subsection{Pattern function reconstruction of the density matrix}\label{PatFunc}

    In the previous section, we studied the phase coherence of a polariton state via phase space methods including Glauber-Sudarshan quasiprobabilities with equivalent measures in the experiment.
    In this section, we discuss an alternative approach to evaluate quantum coherence by explicitly reconstructing the density matrix \cite{pukropDiss}.
    This renders it possible to directly apply the definition of quantum coherence as the sum over the off-diagonal elements of the density matrix as stated in Eq. (\ref{eq:CohNorm}).
    We explicitly reconstruct the density operator in Fock space with the help of so-called pattern functions \cite{PhysRevA.52.R1801,PhysRevA.52.4899,RICHTER1996327,LEONHARDT1996144} by sampling the phase space spanned by the phase space variables $\alpha$.
    The variables individually correspond to an arbitrary single-mode ladder operator $\hat{a}$ of the quantum state \cite{pukropDiss}.
    Extensive reviews are given in Refs. \cite{RevModPhys.81.299,WELSCH199963}.

    In the present work, the Wigner function is sampled numerically via the truncated Wigner approximation.
    The corresponding pattern function can be derived by utilizing the existing mappings between the density matrix $\hat{\rho}$ and the Wigner function $W(\alpha)$, allowing reciprocal reconstruction \cite{PhysRev.177.1882}
    \begin{equation}
    \label{eqn:rhoW}
        \hat{\rho}=\int d^2\alpha W(\alpha)2\hat{\Pi}(\alpha),
    \end{equation}
    and
    \begin{equation}
    \label{eqn:Wrho}
        W(\alpha)=tr\left[\hat{\rho}2\hat{\Pi}\right],
    \end{equation}
    with the displaced parity operator $\hat{\Pi}(\alpha)=\hat{D}(\alpha)\hat{\mathcal{P}}\hat{D}^{\dagger}(\alpha)=(-1)^{\hat{n}(\alpha)}$, where $\hat{D}(\alpha)=\textrm{exp}(\alpha\hat{a}^{\dagger}-\alpha^{\ast}\hat{a})$ is the displacement and $\hat{\mathcal{P}}=\sum_{n=0}^{\infty}(-1)^n\ket{n}\bra{n}=\textrm{exp}(i\pi\hat{n})=(-1)^{\hat{n}}$ the parity operator \cite{pukropDiss}.
    In Eq. (\ref{eqn:rhoW}), the Wigner functions serve as weights for the displaced parity operators representing an expansion of the density matrix.
    By using the link of the trace operation between the expectation value of an arbitrary observable $\hat{L}=\ket{m}\bra{n}$ and the density matrix
    \begin{equation}
        \braket{\hat{L}}=\textrm{tr}\left[\hat{\rho}\hat{L}\right]=\sum_k\braket{k|\hat{\rho}|m}\braket{n|k}
    \end{equation}
    and applying this to Eq. (\ref{eqn:rhoW}) yields
    \begin{equation}
        \braket{n|\hat{\rho}|m}=\int d^2\alpha W(\alpha)2\braket{n|\hat{\pi}(\alpha)|m}=\int d^2 \alpha W(\alpha)2\braket{n|(-1)^{\hat{n}(\alpha)}|m}. 
    \end{equation}
    The pattern functions corresponding to the Wigner function can then be defined as
    \begin{equation}
    \label{eqn:fnm}
        f_{n,m}(\alpha)\equiv 2\braket{n|(-1)^{\hat{n}(\alpha)}|m}.
    \end{equation}
    Within the truncated Wigner approximation we numerically sampling the Wigner function over $N_s$ samples.
    The expectation values of the observable $\hat{L}$ then give the averaged density matrix elements
    \begin{equation}
        \braket{n|\hat{\rho}|m}\approx \frac{1}{N_s}\sum_{i=1}^{N_s}f_{n,m}(\alpha_i).
    \end{equation}
    For our specific system of polariton condensates, any single mode in real or momentum space can be used for the variable $\alpha_i$.
    In the following, we use the representation of number states for the explicit evaluation of the pattern functions
    \begin{equation}
        \ket{m}=\frac{(\hat{a}^{\dagger})^m}{\sqrt{m!}}\ket{0}=\frac{1}{\sqrt{m!}}\partial^m_{\gamma}e^{\gamma\hat{a}^{\dagger}}\ket{0}\bigg \vert_{\gamma=0}=\frac{1}{\sqrt{m!}}\partial^m_{\gamma}e^{|\gamma|^2/2}\hat{D}(\gamma)\ket{0}\bigg \vert_{\gamma=0}=\frac{1}{\sqrt{m!}}\partial^m_{\gamma}e^{|\gamma|^2/2}\ket{\gamma}\bigg\vert_{\gamma=0}.    
    \end{equation}
    In the same manner, we have
    \begin{equation}
        \bra{n}=\frac{1}{\sqrt{n!}}\partial^n_{\beta^{\ast}}e^{|\beta|^2/2}\bra{\beta}\bigg\vert_{\beta^{\ast}=0},
    \end{equation}
    where the displacement operator is used in the representation $\hat{D}(\gamma)=e^{-|\gamma|^2/2}e^{\gamma\hat{a}^{\dagger}e^{\gamma^{\ast}}\hat{a}}$ acting as $\hat{D}(\gamma)\ket{0}=\ket{\gamma}$.
    Additionally, the parity operator in normal order reads \cite{PhysRevA.89.043829}
    \begin{equation}
    (-1)^{\hat{n}}=:e^{-2\hat{n}}.
    \end{equation}
    Applying this expression gives
    \begin{equation}
        (-1)^{\hat{n}(\alpha)}=\hat{D}(\alpha)(-1)^{\hat{n}}\hat{D}^{\dagger}(\alpha)=\hat{D}(\alpha):e^{-2\hat{n}}:\hat{D}^{\dagger}(\alpha)=:e^{-2(\hat{a}^{\dagger}-\alpha^{\ast})(\hat{a}-\alpha)}:
    \end{equation}
    for the displaced parity operator which allows to substituting these expressions back into Eq. (\ref{eqn:fnm}), giving an explicit formula for the pattern functions
    \begin{equation}
    \label{eq:patternf}
    \begin{split}
        f_{n,m}(\alpha)&=\frac{2e^{-2|\alpha|^2}}{\sqrt{n!m!}}\partial^m_{\gamma}\partial^n_{\beta^{\ast}}e^{|\beta|^2/2+|\gamma|^2/2}\bra{\beta}:e^{-2(\hat{a}^{\dagger}-\alpha^{\ast}(\hat{a}-\alpha))}:\ket{\gamma}\bigg\vert_{\gamma=0=\beta^{\ast}}\\
        &=\frac{2e^{-2|\alpha|^2}}{\sqrt{n!m!}}\partial^m_\gamma\partial^n_{\beta^{\ast}}e^{|\beta|^2/2+|\gamma|^2/2}e^{-2(\beta^{\ast}-\alpha^\ast)(\gamma-\alpha)e^{-|\beta|^2/2-|\gamma|^2/2+\beta^\ast\gamma}}\bigg\vert_{\gamma=0=\beta^\ast}\\
        &=\frac{2e^{-2|\alpha|^2}}{\sqrt{n!m!}}\partial^m_{\gamma}\partial^n_{\beta^{\ast}}e^{2\alpha\beta^{\ast}+2\alpha^{\ast}\gamma-\beta^{\ast}\gamma}.
    \end{split}
    \end{equation}
    A known numerical issue when evaluating such pattern functions is the problem of large number calculation.
    This is a common issue that has already been discussed in earlier works \cite{PhysRevA.52.4899} and can be solved by symbolic computation increasing the numerical effort significantly but giving accurate results.

\begin{figure}
    \includegraphics[scale=0.43]{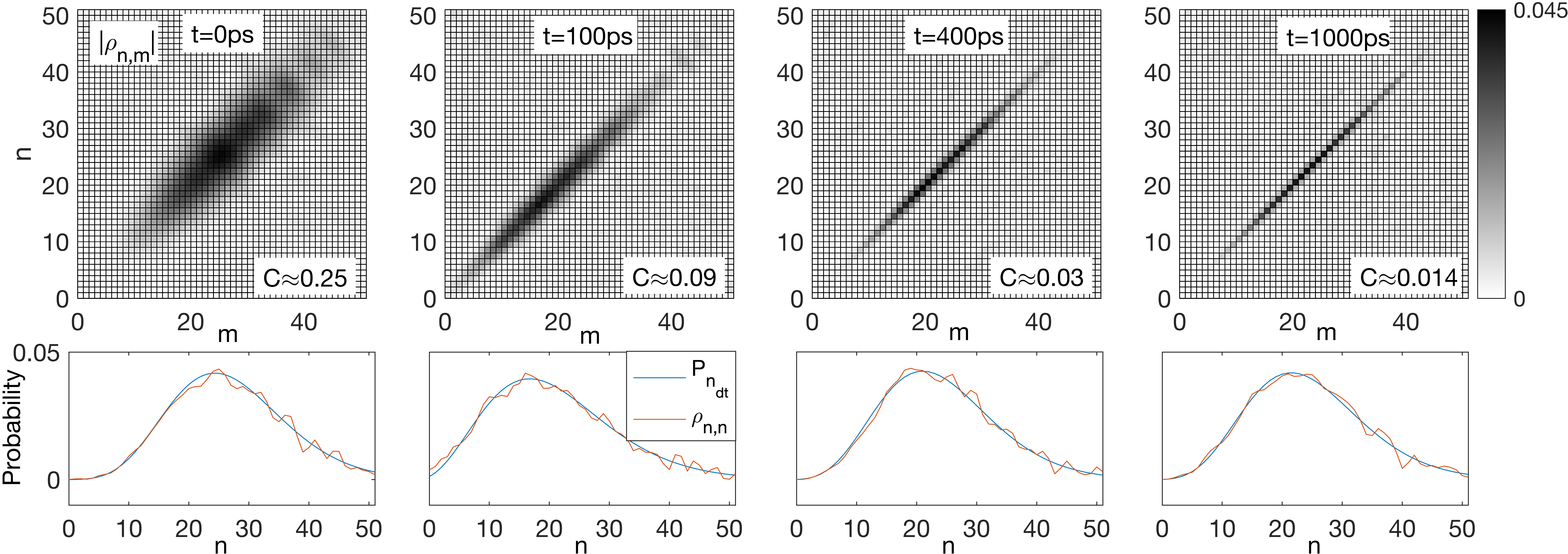}
    \caption{
        Time evolution of the reconstructed density matrices $|\rho_{\textrm{n,m}}|$ (upper row) and photon number distributions (red) compared to the photon statistics of a displaced thermal state (blue) (lower row) corresponding to $P_{\Omega}$ in Fig. \ref{fig:numP}.
        Over time, the off-diagonal elements are contracted towards the diagonal, showing the reduction of quantum coherence.
    }\label{fig:numDensity}
\end{figure}

    We apply the reconstruction method to our simulated data and study the temporal decay of quantum coherence at different points in time.
    We reconstruct the density matrix for the elements $n, m\in[0,51]$, truncating the Fock space as shown in Fig. \ref{fig:numDensity}.

    The diagonal elements yield the occupation number distribution of the state $P_n=\textrm{diag}[\rho_{n,m}]$ and are also shown in Fig. \ref{fig:numDensity}.
    The result can be compared to the theoretical occupation number distribution $ P_{n_{\mathrm{dt}}}$ of a displaced thermal state \cite{1074084}
    \begin{equation}
        \label{eq:Pn}
        P_{n_{\mathrm{dt}}}=\frac{\bar{n}^n}{(1+\bar{n})^{n+1}}\textrm{exp}\left(-\frac{|\alpha_0|^2}{1+\bar{n}}\right)L_n\left(-\frac{|\alpha_0|^2}{\bar{n}+\bar{n}^2})\right),
    \end{equation}
    where $L_n$ are the Laguerre polynomials of $n$-th order.
    The contributions $|\alpha_0|^2$ and $\bar{n}$ are calculated according to the relations given in Eqs. (\ref{eq:nc}) and (\ref{eq:var}).
    Figure \ref{fig:numDensity} features the occupation number distribution of the initial state at $t=0$~ps evaluated via Eq. (\ref{eq:Pn}) in blue and the pattern functions for the zero-momentum mode of the simulated polariton condensate in red.
    The number of simulated realizations is $N_s=4500$, and the condition $\textrm{tr}(\rho)=1$ is fulfilled.
    The reconstructed distribution almost perfectly fits to the values of $P_{n_{\mathrm{dt}}}$ evaluated via Eq. (\ref{eq:Pn}) and is especially well resolved for low occupation numbers.
    The convergence of the reconstructed density matrix elements directly depends on the occupation numbers, the value of $\alpha$, and the number of realizations, and therefore happens asymmetrically.
    Hence, for higher occupation numbers, for the large albeit finite number of realizations, full numerical convergence is not yet reached and the analytical distribution is well but not perfectly reproduced.
    Comparisons with Poissonian and Bose-Einstein distributions show no accordance with the simulated data.
    This further supports that our system just above the excitation threshold is indeed in the transition regime between thermal and coherent emission resulting in a displaced thermal state. 
    We note that at $t=100~$ps in Fig.~\ref{fig:numDensity} a slight shift of the maximum of the photon statistics to lower occupation numbers is observed.
    This is a result of a small deviation of the numerically prepared initial condition from the ideal steady-state solution as discussed above.
    We should emphasize here that the photon statistics alone cannot provide unambiguous state characterization.
    Further information on the state is included in the $P_{\Omega}$ function or the full density matrix including off-diagonal elements.
    While the time evolution of the photon statistics shows that the photon statistics do not change significantly over time in Fig. \ref{fig:numDensity}, $P_{\Omega}$ is annulus shaped revealing additional information and indicating not only a displaced thermal but a decoherent displaced thermal state at $t=1000$~ps.

    The photon-number distribution emitted from an exciton-polariton condensate confined in a micropillar cavity has been measured experimentally by Klaas et al. in Ref. \cite{PhysRevLett.121.047401}.
    In their work, the experiment evaluates the photon number distributions for different excitation powers above threshold and measures displaced thermal states using a photon number resolving transition-edge sensor.
    The results are verified by comparison with the theory evaluated via Eq. (\ref{eq:Pn}).
    For very high excitation powers, the photon-number distributions shift to quasi-Poissonian distributions, demonstrating the transition from a displaced thermal state to coherent emission. 
    Importantly, our numerical results fit nicely these experimental results.
    Still, the photon statistics do not provide any phase information of the state, and in general, the photon statistics cannot give unambiguous state descriptions.
    We are able to further evaluate the phase information of the state by means of the regularized $P$ function, shown in Fig. \ref{fig:numP}.
    Moreover, the measurements give no insight into the off-diagonal contributions of the density matrix.
    Here, we provide numerically calculated off-diagonal elements of the reconstructed density matrix of a polariton condensate.

    The absolute values of the reconstructed density matrix elements of the initial state at $t=0$~ps are given in Fig. \ref{fig:numDensity}.
    The initial state shows a broad elliptical distribution with maximum occupation number at $n=22$. 
    The quantum coherence corresponding to the state can then be calculated by direct summation of the squared absolute values of the off-diagonal elements of the reduced density matrix.
    This yields a quantum coherence of $\mathcal C\approx 0.25$ which is in good agreement with the target value of $\mathcal C=0.24$ calculated via Eq. (\ref{eq:Coherence}).
    Deviations are due to the truncation of the Fock space and the number of realizations. 

    We demonstrate the tracking of quantum coherence using the reconstruction method for the state for the example of a pump intensity of $P=1.25P_{\textrm{th}}$.
    To study the temporal decay of the quantum coherence, the density matrix elements are reconstructed at different points in time.
    In Fig.~\ref{fig:numDensity} the reconstructed density matrices are shown for the times $t=0$~ps, $t=100$~ps, $t=400$~ps, and $t=1000$~ps.
    After 100~ps, the initially broad ellipse in $|\rho_{n,m}|$ is significantly contracted towards the diagonal, indicating the decrease of off-diagonal elements.
    Summation of the absolute value of the off-diagonal elements yields a reduced quantum coherence of $\mathcal C\approx0.09$.
    This trend continues as time progresses with $\mathcal C=0.03$ at $T=400$~ps, and $\mathcal C=0.014$ at $T=1000$~ps.
    At this time, almost no quantum coherence can be detected.    

    It should be noted that for evaluating the quantum coherence in time, the reconstruction method is advantageous compared to the evaluation via Eq. (\ref{eq:Coherence}) as it not only provides more information, but also since Eq. (\ref{eq:Coherence}) is only applicable to displaced thermal states with Gaussian phase space distributions.
    Still, the explicit calculation of the quantum coherence of a state with Gaussian phase-space distribution needs less numerical effort.

    In conclusion, the occupation-number distribution and temporal decay of quantum coherence of polaritonic condensates can be successfully reconstructed via the pattern function approach. 

\section{Summary}

    We presented a detailed discussion on continuous-variable methods in semiconductor spectroscopy.
    We showed how multiport homodyne detection provides detailed insight into the light-matter interaction in semiconductors on ultrafast timescales.
    First, we demonstrated that single port homodyne detection enables us to perform routine measurements such as $g^{(2)}$ measurements of partially coherent emitters requiring much shorter integration times compared to standard discrete-variable measurements based on photodiodes and opens up the possibility to perform real-time measurements.
    Second, we showed that, by using dual-channel homodyne detection, we are able to access the quantum coherence generated by a polariton condensate---a prototypical hybrid quantum system---which is of fundamental importance to quantify its resourcefulness in terms of modern quantum protocols.
    The experimental findings are supported by our numerical simulation utilizing the truncated Wigner approximation to quantify the quantum coherence of the polariton condensate.
    Third, we utilized three-port homodyne detection and postselective conditional homodyne detection to monitor the temporal decay of quantum coherence in the polariton condensate.
    This approach revealed surprisingly long quantum coherence times exceeding 1\,ns, which can be traced back to the fact that the local oscillator in homodyne detection acts as a highly selective filter that reduces the detrimental effects of inhomogeneous broadening strongly.
    Similar coherence times were obtained in the numerical simulation.
    Additionally, we further determined the quantum state by numerically reconstructing the density matrix using pattern functions.

    The combination of cutting-edge phase space methods from quantum optics, tailored resource quantifiers from quantum information science and ultrafast semiconductor spectroscopy techniques renders continuous variable spectroscopy a highly interesting multidisciplinary field of research, which will be essential for future applications of ultrafast quantum information science and technology.

\section*{Funding}
    The authors acknowledge funding through the Deutsche Forschungsgemeinschaft (DFG, German Research Foundation) via the transregional collaborative research centers TRR 142 (Projects A04 and C10, Grant No. 231447078) and TRR 160 (Project B07, Grant No. 249492093).
    A grant for computing time at the Paderborn Center for Parallel Computing (PC$^2$) is gratefully acknowledged.
    The work was further funded through the Ministerium f\"ur Kultur und Wissenschaft des Landes Nordrhein-Westfalen, PhoQC initiative.

\section*{Acknowledgments}
    We gratefully acknowledge Christian Schneider and Sven H\"ofling for providing us with the polariton sample. 

\section*{Disclosures}
    The authors declare no conflicts of interest.

\section*{Data availability}
    Data underlying the results presented in this paper are not publicly available at this time but may be obtained from the authors upon reasonable request.


\bibliography{CVBIB}






\end{document}